# A full-time scale energy management and battery size optimization for off-grid renewable power to hydrogen systems: A battery energy storage-based grid-forming case in Inner Mongolian

Jie Zhu

*College of Electrical Engineering, Sichuan University, Chengdu, 610025, China*

**Abstract:** Hydrogen plays an important role in the context of global carbon reduction. For an off-grid renewable power to hydrogen system (OReP2HS), a grid-forming (GFM) source is essential to provide frequency and voltage references. Here, we take battery works as a GFM source, and the OReP2HS we focus on is comprised of solar photovoltaic, wind turbines, and alkaline electrolyzers for hydrogen generation. An elaborative full-time scale energy management strategy (EMS) covers GFM control to system scheduling (from milliseconds to hours) and is proposed to support high-precision production simulations. Loads of electrolysers are designed to track the renewable power closely to partly replace the energy balancing requirement of the battery. A continuous operation strategy during an emergency like a unit drop is also included in the presented EMS. An off-line simulation-based iterative searching algorithm is developed to find the most suitable size of the battery with the minimum levelized cost of hydrogen (LCOH). A practical OReP2H project located in Inner Mongolian, China is taken as a case study. Simulation results demonstrate the feasibility of the proposed method, and the optimized capacity of the battery (3.4MWh) only accounts for 13.6% of the total installed capacity of the sources with the proposed EMS. Sensitivity analysis shows that LOCH rises from 28.829 CYN/kg to 37.814 CYN/kg, with the time-step for fast power regulation of electrolysers changing from 4s to 1 min with a ramp rate of 0.05MW/s.

| **Nomenclature** | | | $\Delta t_{\text{ROSE}}$ | Time step of the scheduling |
| --- | --- | --- | --- | --- |
| | | | $\Delta t_{\text{SLF}}$ | Time step of instruction issue of SLF control |
| *Abbreviations* | | | $\Delta E_j^{\text{degra,max}}$ | The maximum degradation of a healthy battery |
| | | | $\alpha$ | Weigh of the MA |
| AC | Alternating current | | $\beta$ | Intercept of the SOC linear rectify control |
| BHRES | Battery-hydrogen-based renewable energy system | | $\lambda_j^{\text{O\&M}}$ | The fixed ratio of the annual O&M cost to the initial investment cost |
| CODE | Continuous operation during emergency | | | |
| EMS | Energy management strategy | | $\eta^{\text{Bat}}$ | Charging/discharging efficiency of battery |
| GFM | Grid-forming | | | |
| HESS | Hydrogen-based energy storage system | | *Variables* | |
| LCOE | Levelized cost of energy | | | |
| LCOH | Levelized cost of hydrogen | | $C_{\text{inve}}$ | Initial investment cost |
| MA | Moving average | | $C_{\text{O\&M}}^{\text{fixed}}$ | Fixed annual O&M cost |
| MILP | Mixed-integer linear programming | | $C_j^{\text{rep}}$ | Battery replacement cost |
| MPPT | Maximum power point tracking | | $C_j^{\text{rec}}$ | Battery recycling revenue |
| OReP2HS | Off-grid renewable power-to-hydrogen system | | $C_{\text{O\&M}}^{\text{vari}}$ | Variable annual O&M cost |
| PCC | Point of common connection | | $iter$ | Iterations of the battery size optimization |
| PMEFC | Proton exchange membrane fuel cell | | $M^{H_2}$ | Annual hydrogen yield |
| PSO | Particle swarm optimization | | $K_j^{\text{rep}}$ | Battery replacement times |
| PV | Photovoltaic | | $P_{i,t}^{\text{AE}}$ | Power of AE |
| ReP2H | Renewable power to hydrogen | | $P_{\text{sb}}^{\text{AE}}$ | Standby power of AE |
| RoCoF | Rate of change of frequency | | $P_{i,t}^{\text{AE,ROS}}$ | Baseline load power command of AE |
| ROSEs | Rolling optimal scheduling of electrolysers | | $P_{i,t}^{\text{AE,SLF}}$ | Load power correction command of AE |
| SLF | Self-driven load following | | $P_t^{\text{Bat,C}}/P_t^{\text{Bat,D}}$ | Charging/discharging power of battery |
| SOC | State of charge of battery | | $\hat{P}_t^{\text{RES}}$ | Fast prediction power with SLF control |
| SoCODE | Strategy of continuous operation during an emergency | | $\hat{P}_{l,t}^{\text{WT,cut}}/\hat{P}_t^{\text{PV,cut}}$ | Power curtailment of WT/solar PV |
| SLF | Self-driven load following | | $\hat{P}_{l,t}^{\text{WT}}/\hat{P}_t^{\text{PV}}$ | Ultra-short-term forecast power of WT/solar PV |
| WT | Wind turbine | | $q_{i,t}^{H_2}$ | Hydrogen production rate |
| | | | $S_j$ | Capacity of device $j$ |
| *Parameters* | | | $S_j^{\text{rep}}$ | Replacement Capacity of device $j$ |
| | | | $S_j^{\text{rec}}$ | Recycling Capacity of device $j$ |
| $C_{\text{up,h}}^{\text{AE}}/C_{\text{up,c}}^{\text{AE}}$ | Operational costs for hot/cold start-up of AE | | $SOC_t^{\text{Bat}}$ | Battery state of charge |
| $C_{\text{down}}^{\text{AE}}$ | Operational costs for shut-down of AE | | $S_i^{\text{AE}}$ | Capacity of AE |
| $C^{H_2}$ | Selling price of hydrogen | | $S_{\text{BT}}^{\text{ds,iter}}$ | Iterative index of the battery size optimization |
| $c_j^{\text{rep}}$ | Unit replacement cost of device $j$ | | $T_j^{\text{Fail}}$ | Durable years of device $j$ |
| $c_j^{\text{rec}}$ | Unit recycling revenue of device $j$ | | $\mu_t^{\text{Bat,C}}/\mu_t^{\text{Bat,D}}$ | Charging/discharging state of battery |

| | | | |
|---|---|---|---|
| $c_j^{\text{unit}}$ | The unit cost of device $j$ | $\mu_{i,t}^{\text{st}}/\mu_{i,t}^{\text{sb}}/\mu_{i,t}^{\text{sp}}$ | Binary variables with a value of 1 indicating the start-up/standby/shut-down state of AE |
| $f_{min}^{\text{PCC}}/f_{max}^{\text{PCC}}$ | The allowable minimum and maximum frequency | | |
| $k_i^{H_2}$ | Fixed energy consumption coefficient for hydrogen | $\mu_{i,t}^{\text{up,h}}/\mu_{i,t}^{\text{up,c}}/\mu_{i,t}^{\text{down}}$ | Binary variables with a value of 1 denotes the actions of hot start-up/cold start-up/shut-down of AE |
| $r$ | Discount rate | | |
| $S_{\text{init}}^{\text{Bat}}$ | Initial size of battery in the Iterative optimization | $\Delta E_j^{\text{degra,year}}$ | Annual degradation of battery |
| $SOC_{min}^{\text{Bat}}/SOC_{max}^{\text{Bat}}$ | The minimum and maximum battery state of charge | *Subscripts* | |
| $s$ | Penalty for power curtailment | | |
| $T_{\text{ROSE}}$ | Total time periods of scheduling | | |
| $T_{min}^{\text{down}}$ | The minimum downtime limit for AE | $t$ | Indicator of time intervals |
| $T_j^{\text{LCC}}$ | Lifetime of project | $i$ | Index of the subscript of aes |
| $V_{min}^{\text{PCC}}/V_{max}^{\text{PCC}}$ | The allowable minimum and maximum voltage | $l$ | Index of the subscript of wts |
| $r_{min}^{\text{AE}}/r_{max}^{\text{AE}}$ | The allowable minimum and maximum load regulation percentage of AE | $j$ | Index of the subscript of devices |
| $\Delta S^{\text{Bat}}$ | Increment of battery capacity updated in each iteration | | |

# 1. Introduction

*1.1 Background and motivation*

In the context of global carbon reduction, renewable power-to-hydrogen (ReP2H) technology is rapidly promoted. Green hydrogen production capacity expands over the world, and it offers a novel way for the integrating consumption of large-scale renewable energy. According to the "White Paper on China's Hydrogen Energy and Fuel Cell Industry", as of 2022, the installed capacity of power sources for hydrogen production has reached 580 MW. Projections indicate that by 2060, the total installed capacity of electrolysers in China will reach 500 GW, with 80% of hydrogen demand met through ReP2H [1] [2].

The scaled ReP2H integrated into the power system can be classified into grid-connected and off-grid types. A grid-connected ReP2H system integrates the surplus power into the utility grid but purchases electricity to maintain hydrogen production when a power shortage occurs. On one hand, the sufficient peak-shaving capacity of the utility grid is required in this operation mode. But as the abundance of grid peak-shaving resources decreases, grid operational constraints limit its scalability [3] [4]. On the other hand, it is challenging to guarantee that all the hydrogen produced by a grid-connected ReP2H system is zero-carbon green hydrogen, and it is exclusively powered by renewable energy, due to the exchange of electricity with the external grid. Moreover, remote areas with abundant renewable resources such as deserts, plateaus, et al. are inadequate coverage of the utility grid.

The off-grid ReP2H system can mitigate these issues caused by electricity exchange with external grids. It can be flexibly designed based on the spatial distribution of renewable resources, and the stringent constraints and regulations on power exchange while integrating into the grid are also avoided. Beneficial from this, both the planning and operational design of an off-grid ReP2H system are convenient, making it the most promising development mode. Meanwhile, the off-grid way has garnered support from various policies, and some off-grid ReP2H demonstration projects are gradually being constructed [5] [6] [7].

To utilize the on-shore wind turbines (WTs) and solar photovoltaic (PV), most off-grid ReP2H projects tend to adopt an alternating current (AC) approach. However, when operated in the AC off-grid mode, a grid-forming (GFM) source works as a controlled voltage source is necessary to provide frequency and voltage references for the ReP2H system. It only can be assumed by internal devices such as WT, PV, battery, or electrolyser. However, the development of WT, PV, and electrolyser equipment with a grid-forming ability are still in the experimental verification stage [8] [9] [10]. Taking battery (electrochemical energy storage) as the grid-forming source of an AC off-grid ReP2H system (OReP2HS) seems to be the most feasible solution [11] [12], but several challenges appeared for selecting the optimum battery size and energy management strategy (EMS).

Firstly, there is only one device, the battery, that works as a grid-forming source in the OReP2HS. The battery has to adjust its output to maintain the stability of the system frequency and voltage whenever the system is disturbed. If the power disturbance exceeds the limitations of battery response, OReP2HS loses the frequency reference and is directly destabilized [13] [14]. Thus, for an OReP2HS, the grid-forming battery with a large capacity and high charging/discharging rate can increase the anti-interference ability of the system, but the investment cost is also increased. A trade-off between the economy of hydrogen production and the stability of system operation becomes the key to system design. Secondly, loads of electrolysers are demonstrated to have a wide range and low-latency response capability. How to fully utilize its superior power regulation performance to replace part of the energy balance regulation from the GFM battery, so as to reduce the battery size and the levelized cost of hydrogen (LCOH) is also an urgent issue to be addressed.

Consequently, to aforementioned challenges leave the issue open to designing EMS and identifying a cost-effective battery size for an OReP2HS. Our research proposes a comprehensive full-time scale EMS for off-grid ReP2H systems, and in combination with the high-precision model and data, the battery size is optimized, while the system stability verification and the energy balance constraints are both considered.

*1.2 Literature review*

Currently, the designing, planning, and operational scheduling issues of an OReP2HS have received extensive attention by scholars worldwide.

Most of them regard power-to-hydrogen as an energy storage medium to enable energy independence in an off-grid system. Marocco et al. [15] [16] [17] design a battery-hydrogen-based renewable energy system (BHRES) with H2 and batteries as hybrid energy storage to keep the energy autonomous in the insular or remote areas. The BHRES is comprised of PV/WT, batteries, a hydrogen-based energy storage system (HESS) consisting of an electrolyser, a fuel cell, and pressurized hybrid tanks. The components capacities of the BHRES are optimized by the particle swarm optimization (PSO) with 1h time resolution based on a rule-based EMS proposed in [16]. A techno-economic analysis considering the degradation cost of batteries and the fuel cell is implemented in [17]. It indicates that the levelized cost of energy (LCOE) of BHRES increases from 0.455 €/kWh to 0.544 €/kWh with only batteries as storage medium. Gandiglio et al. [18] sequentially carry out a life cycle assessment of the BHRES to evaluate the environmental improvement compared to the diesel generator-based configuration. The results show that using power-to-hydrogen to smooth the solar power fluctuations, leads almost 10% reduction in climate change, ozone depletion, acidification, and other environmental indicators presented in the paper. Babaei et al. [19] utilize the BHRES to supply power for the energy-stressed islands in Eastern Canada, and the HOMER software is employed to find the most cost-effective configuration. Also, the BHRES plays an important role in achieving energy independence for green buildings [20] [21], standalone communities [22] [23] [24] and villages [25], off-grid industrial parks [26], and research facility in isolated regions [27].

Part of the researchers replaces the electrolyser and fuel cell in BHRES by a proton exchange membrane fuel cell (PMEFC) device. Zhang et al. [28] analyze the techno-economic feasibility of developing a zero carbon emission aim in remote islands under different combinations of renewable energies (PV/WT/Hydro generator) with PMEFC. A practical case in the Ui island of South Korea is studied, and the LOCE is 0.366 $/kWh with the best configuration of PV/WT/battery/PEMFC. Ghenai et al. [29] design an off-grid PV/PEMFC system to satisfy the electric needs of a residential community located in a desert region. The PEMFC worked as an energy storage device and supplied 42% power of the total load demands. Then, Ceylan et al. [30] determine the minimum LCOE of a PEMFC-based hybrid energy system that operates in both off-grid and on-grid modes. Results show that the LCOE increases from 0.223 $/kWh to 0.410 $/kWh when the system switches from grid-connected to off-grid operation.

Some scholars take hydrogen to match the long-term energy balances. Shao et al. [31] consider hydrogen as seasonal energy storage, and a bi-level optimization is proposed to determine the sizing and scheduling of the electrolyser and hydrogen tanks. Nastasi et al. [32] compare the economic performance of the hydrogen and battery-based energy storage options for the independent operation of a community-powered by PV/WT. A simulation-based planning is implemented using the DECAPLAN™ digital platform, which shows that hydrogen-based storage is a more attractive solution. Abdin et al. [33] also suggest that hydrogen has economic benefits over batteries for long-period energy storage in off-grid energy systems.

For green hydrogen production towards transportation or industrial applications, Ibagon et al. [34] investigate the potential of export-oriented green hydrogen production via an off-grid renewable energy plant in four different locations in Uruguay. It is found that the LCOH decreasing from 3.5 $/kg in 2022 to 2.3 $/ kg in 2030, with the maturity of the hydrogen production technology. Janssen et al. [35] quantify the possible reductions of LCOH until 2050 using solar and on- and offshore wind energy in European countries. They project that the LCOH towards 2050 can decrease below 2 €/kg in several countries in several countries, like Denmark and Ireland. Posso et al. [36] estimate of the green hydrogen production potential in the Americas. Here, the case study demonstrates PV to hydrogen is the most economically preferable route. Ibáñez-Rioja [37] [38] develop a rule-based operational strategy to support the optimal planning of an off-grid green hydrogen production plant. It is found that the LCOH can be reduced to 2 €/kg by the year 2030 in southeastern Finland. Wang et al. [39] focused on the green hydrogen production for green ammonia synthesis. Simulations and the optimizations derived into a mixed-integer program form are carried out to identify the cost-effective configurations of OReP2HS. To meet the hydrogen vehicle refueling requirements in Dhahran City, Pang et al. [40] propose a mixed integer quadratic constrained program framework for the optimal designing of the off-grid hybrid energy refueling system. To get highly reliable hydrogen production, the minimum daily refueling cost is 36.32 $/kg. Abdulrahman et al. [41] utilize surplus renewable power to produce hydrogen for refueling hydrogen-based driven vehicles in the off-grid residential area. Through an hourly production simulation, the system reduces the CO2 emission by 9.66 Tons per year. Yang et al. [42] design a hydrogen supply chain network in Fujian province, China and an optimization is conducted to address the planning and operation issues of it. The LCOH was optimized ranging from 3.037 $/kg to 3.155 $/kg. Tang et al. [43] perform a techno-economic feasibility assessment of hydrogen refueling stations, and an off-grid scenario brings better performance with the LOCH ranging from 3.5 €/kg to 7.2 €/kg.

In the literature reviewed above, the planning and scheduling of OReP2HS are mainly implemented with an optimization- or a rule-based production simulation, and its main features are presented in Table 1.

Table. 1 Summary of the key information for the OReP2HS designing of the reviewed literatures.

| Literature | Components | Long-term energy balance | Optimal scheduling | Hydrogen supply reliability | Operation stability | Time resolution | Method/Software |
|---|---|---|---|---|---|---|---|
| [15] [18] | PV/WT/BT[1]/HESS | √ | × | × | × | 1h | / |
| [16] [20] [22] | PV/WT/BT/HESS | √ | × | × | × | 1h | PSO algorithm |
| [17] | PV/WT/BT/HESS | √ | × | × | × | 1h | MILP |
| [19] [23] [25] [28] | PV/WT/BT/HESS | √ | × | × | × | 1h | HOMER |
| [24] | PV/WT/SC[2]/HESS/Micro-hydro power plant | √ | × | × | × | 1h and 1 month | metaheuristic algorithm |
| [26] | PV/WT/HESS | √ | × | × | × | 1h | NSGA-II algorithm |
| [27] | PV/WT/BT/Diesel generator/HESS | √ | × | × | × | 1h | highRES-AtLAST[4] |
| [29] | PV/HESS | √ | × | × | × | 30 mins | Matlab/Simulink |
| [30] | PV/WT/HESS | √ | × | × | × | 1h | Matlab Simulink |
| [31] [32] | PV/WT/BT/HESS | √ | √ | × | × | 1h | MILP |
| [33] | PV/WT/BT/HESS | √ | × | × | × | 1h | HOMER Pro |
| [34] | PV/WT/BT/AE | √ | × | × | × | 1h | Sequential quadratic programming |
| [37] | PV/WT/BT/AE | √ | × | × | × | 5 mins | PSO algorithm |
| [38] | PV/WT/BT/PEM[3] | √ | × | × | × | 5 mins | evolutionary algorithm |
| [39] | PV/WT/BT/AE | √ | √ | × | × | 1h | MILP |
| [40] | PV/WT/BT/AE/dispenser | √ | × | √ | × | 1h | HOMER |
| [41] | PV/WT/BT/AE/dispenser | √ | × | √ | × | 1h | Matlab |
| [42] | PV/WT/BT/AE/ dispenser | √ | × | √ | × | 1h | MILP |
| [43] | PV/WT/AE | √ | × | × | × | 1h | / |

[1] Battery; [2] Super capacitor; [3] Proton exchange membrane electrolyser; [4] An highRES model for the energy system of Great Britain and Europe.

As can be observed, most of the existing studies only focus on the long-term energy balance with a time resolution ranging from 5 minutes to 1 hour. The involved models of renewable sources are derived equivalently, only considering power balance, but ignoring the transient dynamic of the equipment. Also, the operation stability of OReP2HS is not taken into account.

However, without the inertia and energy supported by the utility grid, the voltage and frequency stability of an OReP2HS become non-negligible problems. Specifically, the power sources are equipped with a GFM control in OReP2HS worded like grid-tied synchronous generators. They increase or decrease their output power instantaneously to balance loads and maintain system voltage and frequency when disturbances occur. If the imbalance power is beyond the regulation capacity of GFM sources, the OReP2HS may be directly destabilized or be forced to be shut-down imminently due to the reference loss of the voltage and frequency [44] [45]. The process by which GFM sources respond to disturbances that cause instability is almost milliseconds, and involves complex interactions among multiple power devices. Thus, the production simulation performed with a low time resolution (minutes to an hour) and equivalent models hardly fully reveal the practical operating status of the OReP2HS, leading to an imprecise and idealistic sizing result.

On the other hand, towards transportation or industrial applications, a reliable supply of hydrogen is essential. Few papers take the lack of hydrogen supply probability of the OReP2HS to restrict the shut-down times of electrolyser while conducting the optimal planning. Nevertheless, keeping the continuous green hydrogen production when a power loss occurs like a unit drop is a lack of discussion. During such an emergency, GFM sources need to first provide power support without any time delay to secure the continuity and reliability of the hydrogen supply. Therefore, a capacity margin of GFM sources needs to be allowed for in planning the OReP2HS.

*1.3 Contributions*

In this paper, we present a novel method that optimizes the battery size of an OReP2HS, while considering the long-term energy balance, operation stability, and hydrogen supply reliability. The OReP2HS we focus on consists of solar PV, WTs, a bank of battery works as a GFM source to control the system voltage and frequency, and alkaline electrolysers for hydrogen generation. An elaborative EMS covers from GFM control to system scheduling (from milliseconds to hours) is developed to support high-precision production simulations. Here, loads of electrolysers are designed to track the renewable power to partly replace the energy balancing requirement of the battery. A continuous operation strategy during an emergency like a unit drop is also included in the presented EMS. Simulations performed in this study are based on data collected with 1s~60s time resolution from the practical wind farms and PV stations that are located in Inner Mongolian, China. An iterative searching algorithm is proposed to find the most suitable size of the battery with the minimum LCOH. The potential and limitations of the OReP2HS with the proposed EMS are evaluated through a sensitivity analysis.

The main contributions included in our study are summarized as follows:

- A full-time scale EMS is developed to support high-precision production simulations of the OReP2HS. It covers GFM control to the scheduling of load rotation (from milliseconds to hours) with a three-layer architecture, and it includes a continuous operation strategy addressing the emergency like a unit drop.
- A searching algorithm is proposed to find the most applicable size of the grid-forming battery with the minimum LCOH while considering the constraints of long-term energy balance, GFM stability, and continuous operation ability during an emergency.
- Discussions of the potential and limitations of an OReP2HS with the proposed EMS are evaluated by means of the results of the sensitivity analysis. A practical case study in Inner Mongolian, China shows that LOCH rises from 28.829 CYN/kg to 37.814 CYN/kg, with the time-step for fast power regulation of electrolysers changing from 4s to 1 min with a ramp rate of 0.05MW/s.

The rest of this paper is organized as follows. The configuration of the OReP2HS studied in this paper is presented in Section 2, and the models of the components are described in Section 3. In Section 4, the full-time scale EMS is introduced in detail. Section 5 presents the optimization methods. The results of the simulations and the sensitivity analysis are given in Section 6. Finally, Section 7 presents the conclusion.

## 2. System description

*2.1 Configuration*

A practical OReP2H project located in Inner Mongolian, China is taken as a case study in our research, and its layout is shown in Fig. 1. This project plans to configure three 6.25 MW wind turbines from Goldwind Sci & Tech, and a 6.25 MWp solar PV system from LONGi Green Energy Technology will be installed. Regarding the hydrogen production plant, it is composed of three 5 MW alkaline electrolysers (AEs) from Peric Hydrogen Technologies. Each AE has a rated power of 5 MW and a maximum overload capacity of 120% to rating. The GFM battery plans to be purchased from SUNGROW, but its capacity is still waiting to be identified. All the devices of power sources and hydrogen production are connected by 35kV

AC cables with a rated frequency of 50 Hz. The produced hydrogen is integrated by transport pipelines and fed to the hydrogen consumers after compressed and buffered.

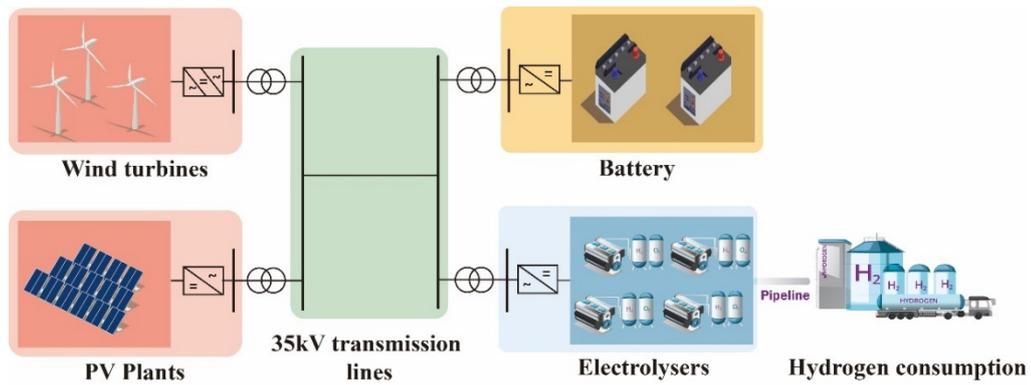

**Fig. 1** Layout of the studied practical OReP2HS located in Inner Mongolia, China.

*2.2 Requirements for the designing*

due to the complexity and energy/time consumption of the black start, it is worthwhile avoiding the unnecessary black stars. Thus, OReP2HS should remain stable under the frequent occurred emergency like a device dropped, to ensure the reliable hydrogen supply and operation. However, at the moment of emergency, OReP2HS only relies on the GFM battery to immediately balance part of the power shortage or surplus. Therefore, the continuous operation ability during an emergency is necessary to incorporate into the identification of the GFM battery.

OReP2HS has to fully utilize its internal GFM battery and electrolysers with rapid load response capabilities to participate in energy balance at multiple time scales, due to the lack of external power grid support. The identification of the GFM battery configuration is crucial for economical, safe, and efficient operation, the constraints the following should be satisfied:

1) **GFM ability**: the GFM battery should provide fast power support to maintain the presence of the frequency and voltage references, whenever and whatever the disturbances occur. It is the primary constraint for the stable operation of an OReP2HS. The GFM ability gets lifted with the increase of battery size and charging/discharging rate, but the investment cost is also increased.

2) **Continuous operation ability during emergency**: due to the complexity and energy/time consumption of the black start, it is worthwhile avoiding the unnecessary black stars. Thus, OReP2HS should remain stable under the frequent occurred emergency like a device dropped, to ensure the reliable hydrogen supply and operation. However, at the moment of emergency, OReP2HS only relies on the GFM battery to immediately balance part of the power shortage or surplus. Therefore, the continuous operation ability during an emergency is necessary to incorporate into the identification of the GFM battery.

3) **Long-term energy balance:** energy balance over a year is a basic constraint in the planning of an OReP2HS. But notably, the AE has the ability to fast regulate its load power, and the regulation range is far beyond that of the battery under the unit combination of AEs. Part of the energy balance regulation for the GFM battery could be replaced by AEs to save the investment cost if the load regulation ability is utilized sufficiently. And this is the core of the proposed full-time scale EMS.

The GFM ability and continuous operation ability during an emergency involves transient processes and the complex interaction between multiple devices, a refined simulation model is necessary for the verification. However, efficiency gets more attention than precision in long-term energy balance checking. Therefore, both the refined and simplified simulation models are developed (detailed introduced in Section 3), to serve the different checking requirements of the aforementioned constraints. Then, the EMS of the OReP2HS on a full-time scale from millisecond to day-ahead is presented (detailed introduced in Section 4), to support the battery size optimization.

## 3. Simulation models

Refined and simplified simulation models are developed the satisfy different checking requirements of the constraints listed in Section 2.

The refined model is developed with electromagnetic transient prediction for checking the GFM ability and continuous operation ability during an emergency. It is established on the Matlab/Simulink software. For the power sources, maximum power point tracking (MPPT) control is integrated into the control loops of the WTs and solar PV, to fully utilize the wind and solar resources. For the load side, the rectifier connected to the electrolyser adopts a dual-stage topology of AC/DC converters in series with boost converters. The equivalent circuit of the electrolyser is developed based on the classic model with internal resistances in series with counter electromotive force to simulate its U-I (voltage-current) feature. The GFM battery is designed based on voltage-frequency (V-F) droop control, and the classic equivalent circuit is employed to simulate the U-I feature of the battery.

The simplified model is established on the foundation of the electromagnetic transient model. However, the switching dynamics within the converters are disregarded, and an ideal voltage source is utilized for modeling the converter. This model aims to improve simulation efficiency while ensuring model accuracy, satisfying the long-term energy balance verification requirement.

Both the refined and simplified simulation models are based on the parameters of actual devices (as introduced in Section 2). A full description of detailed topologies, control loops, and parameters for both models are presented in Appendix A.

## 4. Full-time scale energy management strategy

### 4.1 Framework

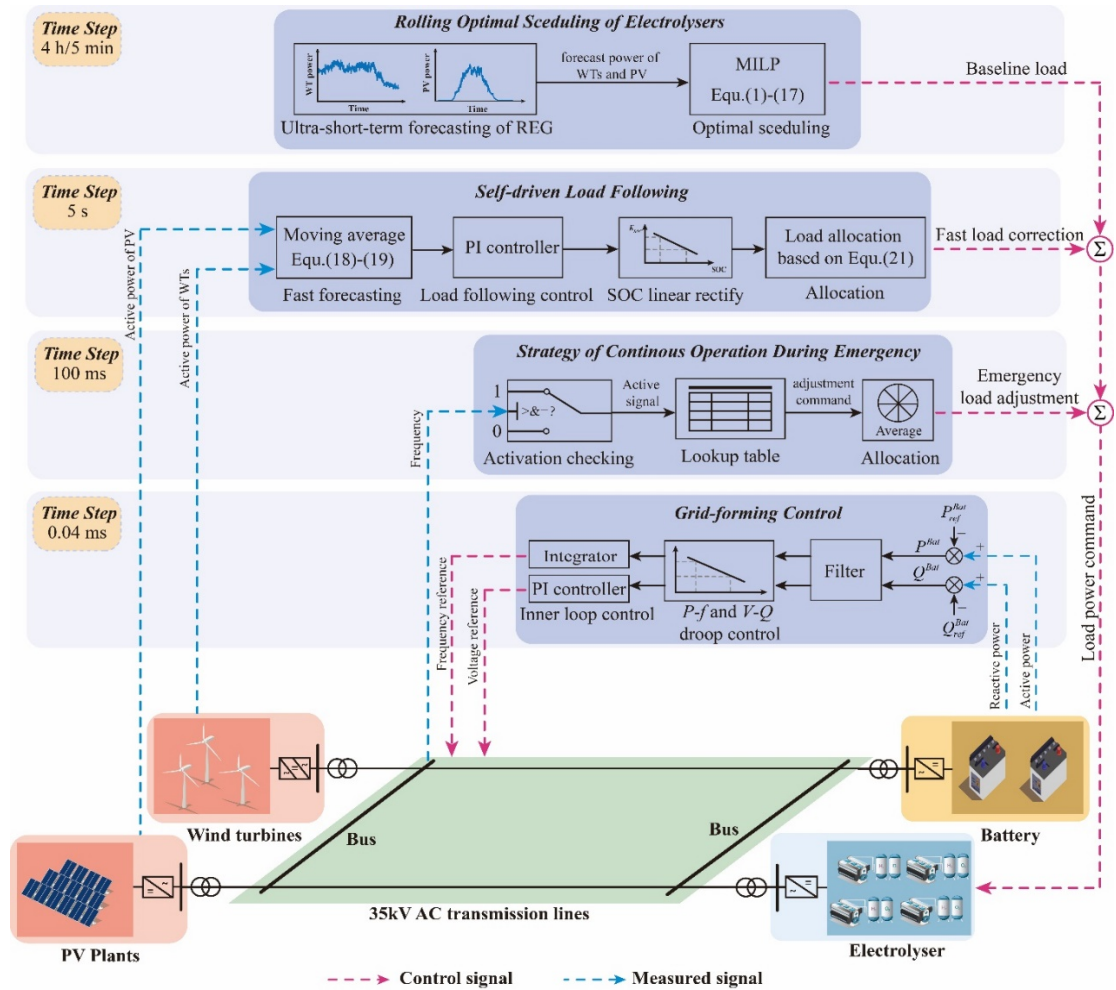

**Fig. 2** Framework of the proposed full-time scale EMS

The framework of the proposed full-time scale EMS is illustrated in Fig. 2. As we can see, WTs and solar PV operate with MPPT controllers. The GFM battery works as a support source, providing the voltage and frequency reference to the OReP2HS. The AEs are crucial in the EMS, they track the renewable power closely with a two-stage load controller, ensuring energy from 5 seconds to 4 hours.

The proposed EMS covers from 0.04 millisecond to 4 hours and four sub-strategies are included: rolling optimal scheduling of electrolysers (5 minutes to 4 hours), self-driven load following strategy (4 seconds), droop-based grid-forming control strategy (millisecond scale), and the strategy of continuous operation during an emergency (0.1 second) is also designed. Based on the forecasts of renewable power for the upcoming 4 hours, the rolling optimal scheduling of electrolysers (ROSEs) utilizes mixed-integer linear programming (MILP) to swiftly determine the start-up/shut-down statuses and baseline power for each AE with a time step of 5 minutes. Self-driven load following (SLF) strategy performs corrections to the loads power of AEs with 5 seconds time resolution, to ensure the loads track the renewable power well. Then, the purpose of releasing the battery regulation demand can be achieved. The GFM control strategy of battery (V/F droop control is adopted in this paper) provides a frequency and voltage reference for the operation of OReP2HS and remains them stable at a typical control time scale of 0.04 millisecond. The strategy of continuous operation during emergency (SoCODE) monitors system frequency continuously, and it is activated to ensure the continuous operation of the OReP2HS in case of an emergency (unit dropped). Four sub-strategies coordinate with each other to achieve a stable and cost-effective operation of the OReP2HS.

*4.2 Rolling optimal scheduling of electrolysers*

The ROSEs swiftly determine the start-up/shut-down statuses and baseline load of each AE for the upcoming 4 hours, according to the ultra-short-term forecasts results of renewable power. A MILP model is developed for the load scheduling of the OReP2HS with a time step of 5 minutes, and the objective is set to minimize the cost of hydrogen production, shown as:

$$\min_{x \in X} \sum_{t=0}^{T_{\text{ROSE}}} \sum_{i=0}^{n} \left\{ (-C^{H_2} q_t^{H_2}) \Delta t_{\text{ROSE}} + C_{\text{up,h}}^{\text{AE}} \mu_{i,t}^{\text{up,h}} + C_{\text{up,c}}^{\text{AE}} \mu_{i,t}^{\text{up,c}} + C_{\text{down}}^{\text{AE}} \mu_{i,t}^{\text{down}} + s \left( \sum_{l=1}^{m} \hat{P}_{l,t}^{\text{WT,cut}} + \hat{P}_t^{\text{PV,cut}} \right) \right\} \quad (1)$$

where, $x$ is the decision variables, encompassing the start-up/shut-down/standby status, and load power of AEs in all the scheduling periods. $T_{\text{ROSE}}$ are the total time periods of scheduling. $C^{H_2}$ is the selling price of hydrogen. $C_{\text{up,h}}^{\text{AE}}$, $C_{\text{up,c}}^{\text{AE}}$, and $C_{\text{down}}^{\text{AE}}$ represent the operational costs for hot start-up, cold start-up, and shut-down of the AE, respectively; $s$ is the penalty for curtailment, and $m$ is the numbers of WTs. $\hat{P}_{l,t}^{\text{WT,cut}}$ and $\hat{P}_t^{\text{PV,cut}}$ respectively indicate the curtailed WT and solar PV power. $\Delta t_{\text{ROSE}}$ is the time step of scheduling, set at 5 minutes. $\mu_{i,t}^{\text{up,h}}$, $\mu_{i,t}^{\text{up,c}}$, and $\mu_{i,t}^{\text{down}}$ are binary variables where a value of 1 denotes the actions of hot start-up, cold start-up, and shut-down of AE.

For the AEs, they can switch their status to start-up, standby, and shut-down flexibly. Notably, in the standby state, only the auxiliary equipment is turned on to maintain the temperature and pressure of the electrolyser, but hydrogen production is stopped. Therefore, the constraints of AEs can be expressed as:

$$\mu_{i,t}^{\text{st}} + \mu_{i,t}^{\text{sb}} + \mu_{i,t}^{\text{sp}} = 1 \quad (2)$$

$$\mu_{i,t}^{\text{down}} = \mu_{i,t-1}^{\text{st}} \mu_{i,t}^{\text{sp}} \quad (3)$$

$$\mu_{i,t}^{\text{up,h}} = \mu_{i,t-1}^{\text{sb}} \mu_{i,t}^{\text{st}} \quad (4)$$

$$\mu_{i,t}^{\text{up,c}} = \mu_{i,t-1}^{\text{sp}} \mu_{i,t}^{\text{st}} \quad (5)$$

$$T_{min}^{\text{down}} \mu_{i,t}^{\text{down}} \leq \sum_{h=t}^{t-1+T_{min}^{\text{down}}} \mu_{i,t}^{\text{sp}}, \quad \forall t \leq T_{\text{ROSE}} + 1 - T_{min}^{\text{down}} \quad (6)$$

$$(1 - \mu_{i,t}^{\text{sb}}) P_{i,t}^{\text{AE,ROS}} = k_i^{H_2} q_{i,t}^{H_2} \quad (7)$$

$$\mu_{i,t}^{\text{sb}} P_{\text{sb}}^{\text{AE}} + \mu_{i,t}^{\text{st}} r_{min}^{\text{AE}} S_i^{\text{AE}} \leq P_{i,t}^{\text{AE,ROS}} \leq \mu_{i,t}^{\text{st}} r_{max}^{\text{AE}} S_i^{\text{AE}} + \mu_{i,t}^{\text{sb}} P_{\text{sb}}^{\text{AE}} \quad (8)$$

(3)-(5) describe the switching of different statue of AEs, where, $\mu_{i,t}^{\text{st}}$, $\mu_{i,t}^{\text{sb}}$, and $\mu_{i,t}^{\text{sp}}$ are binary variables with a value of 1 indicating the start-up, standby, and shut-down state of AEs, respectively. (6) describes the minimum downtime limit for AEs, where $T_{min}^{\text{down}}$ is set at 1 hour. (7) represents the relationship between the load power $P_{i,t}^{\text{AE}}$ of AEs and hydrogen production rate $q_{i,t}^{H_2}$ using a fixed energy consumption coefficient $k_i^{H_2}$. The variation range of load power is presented in (8), where $P_{\text{sb}}^{\text{AE}}$ is the power consumption in standby state, $S_i^{\text{AE}}$ is the capacity of a single AE. $r_{min}^{\text{AE}}$ and $r_{max}^{\text{AE}}$ limit the boundaries of load regulation, typically with $r_{min}^{AE}$=10% and $r_{max}^{AE}$=120%.

Notably, the fixed energy consumption coefficient $k_i^{H_2}$ is only used to simplify the ROSEs to quickly obtain the baseline load power references. A practical nonlinear efficient curve of AE is employed for the LCOH calculation and battery size optimization in Section 5.

The constrains for the battery are listed as follows:

$$SOC_t^{Bat} = SOC_0^{Bat} + \sum_{i=1}^{T_{ROSE}} (\eta^{Bat} P_t^{Bat,C} - \frac{P_t^{Bat,D}}{\eta^{Bat}}) \Delta t_{ROSE} \tag{9}$$

$$|SOC_t^{Bat} - SOC_0^{Bat}| \leq 0.05, \quad t = T_{ROSE} \tag{10}$$

$$SOC_{min}^{Bat} \leq SOC_t^{Bat} \leq SOC_{max}^{Bat} \tag{11}$$

$$0 \leq P_t^{Bat,C} \leq \mu_t^{Bat,C} P_t^{Bat,Cmax} \tag{12}$$

$$0 \leq P_t^{Bat,D} \leq \mu_t^{Bat,D} P_t^{Bat,Dmax} \tag{13}$$

$$0 \leq \mu_t^{Bat,C} + \mu_t^{Bat,D} \leq 1 \tag{14}$$

(9) describes the dynamic change in the state of charge (SOC) of battery, where, $\eta^{Bat}$ is charging/discharging efficiency. $P_t^{Bat,C}$、$P_t^{Bat,D}$ represent the charging and discharging power, respectively. (10) indicates the cycle equilibrium constraint. The maximum charging and discharging limits are described in (11). (12)-(14) restrict the battery working at a charging and discharging simultaneously.

The constrains of power balance is written as:

$$\sum_{l=1}^{m} \hat{P}_{l,t}^{WT} + \hat{P}_t^{PV} - \sum_{l=1}^{m} \hat{P}_{l,t}^{WT,cut} + \hat{P}_t^{PV,cut} + P_t^{Bat,D} = P_t^{Bat,C} + \sum_{i=1}^{n} P_{i,t}^{AE,,ROS} \tag{15}$$

$$0 \leq \hat{P}_{l,t}^{WT,cut} \leq \hat{P}_{l,t}^{WT} \tag{16}$$

$$0 \leq \hat{P}_t^{PV,cut} \leq \hat{P}_t^{PV} \tag{17}$$

where, $\hat{P}_{l,t}^{WT,cut}$ and $\hat{P}_t^{PV,cut}$ are the power curtailment of WT and solar PV, respectively.

Finally, the proposed ROSEs model can be summarized as (1)-(17) a framework of MILP. Thus, it can be easily solved by an off-the-shelf commercial solver.

*4.3 Self-driven load following*

SLF strategy provides a correction for the load power of AEs. It aims to fully use the advantages of the fast load regulation capacity of AEs to smooth the unbalanced power within 5 seconds. Four sub-modules: load following control, SOC linear rectification, and allocation are integrated into the SLF strategy, as illustrated in Fig. 1.

In terms of fast forecasting, a fast power prediction model is established based on the discrete moving average (MA) process, as shown in (18). The operation monitoring data of WTs and solar PV outputs are employed, and the correction command of the load power is calculated by (19).

$$\hat{P}_t^{RES} = \alpha \hat{P}_{t-\Delta t_{SLF}}^{RES} + (1-\alpha) \frac{1}{q_{SLF}} \sum_{k=1}^{q_{SLF}} \left( \sum_{l=1}^{m} P_{l,t-k\Delta t_{SLF}}^{WT} + P_{t-k\Delta t_{SLF}}^{PV} \right) \tag{18}$$

$$P_t^{SLF} = \hat{P}_t^{RES} - \sum_{i=1}^{n} P_{i,t-\Delta t_{SLF}}^{AE} \tag{19}$$

where, $\hat{P}_t^{RES}$ is the fast prediction power, $q_{SLF}$ is order of the MA, and it is set at 4. $\Delta t_{SLF}$ indicates the time step of the SLF.

The PI controller is employed to track the correction command of the load power. At the end of this controller, a SOC linear rectify control module is connected, to avoid the energy waste of the battery caused by the over-charging or over-discharging. Thus, the calculation of the total correction command can be written as:

$$P_t^{SLF,ref} = K_{SOC}(K_P P_t^{SLF} + \int K_I P_t^{SLF}) + \beta \tag{20}$$

where, the $K_{SOC}$ and $\beta$ are the coefficients of the SOC linear rectify control module.

The available space for load correction of each AE may be different due to the difference in their baseline power references. Therefore, an allocation method for distributing the total correction command to each AE based on their adjustable capacity is presented in (21).

$$P_{i,t}^{AE,SLF} = \begin{cases} \frac{S_i^{AE} - P_{i,t}^{AE}}{\sum_{i-1}^{n}(S_i^{AE} - P_{i,t}^{AE})} P_t^{SLF,ref}, & \text{for upward regulation} \\ \frac{P_{i,t}^{AE}}{\sum_{i-1}^{n} P_{i,t}^{AE}} P_t^{SLF,ref}, & \text{for downward regulation} \end{cases} \tag{21}$$

Finally, the load power reference of AEs can be written as:

$$P_{i,t}^{AE} = P_{i,t}^{AE,ROS} + P_{i,t}^{AE,SLF} \tag{22}$$

where, $P_{i,t}^{AE,ROS}$ and $P_{i,t}^{AE,SLF}$ are the baseline and correction load commands, and they are updated every 5 seconds and 5 minutes, respectively.

*4.4 Strategy of continuous operation during an emergency*

SoCODE is proposed for preventing the unnecessary black star of OReP2HS under the frequently occurring emergency as a unit dropped. It is developed based on the strong coupling relationship between frequency and active power of the under V/F droop control of the GFM battery. The rate of change of frequency (RoCoF) and the maximum frequency deviation are simultaneously used for judging the emergency severity of an accident. Once the SoCODE is activated, a fast power regulation command issues to the generations or the loads (depending on wherever the unit drop occurred on the load side or the sources side) based on the lookup table (as shown in Table 2), to ensure the power of OReP2HS quickly goes back to balance. Thus, the power support demand from the battery is released. Also, the size and the investment cost of the battery can be saved.

Notably, if the indicators of RoCoF or the maximum frequency deviation do not simultaneously fall within the interval listed in Table 2, SoCODE cannot be activated. The frequency is regulated in an allowable interval by the p-f droop control of the GFM battery when the SoCODE is quit.

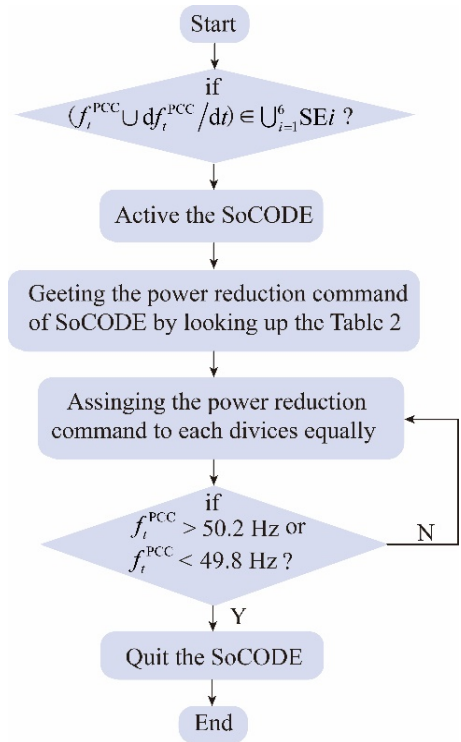

**Fig. 3** Flow chart of the excitation of the SoCODE.

**Table. 3** Power reduction and active requests of the SoCODE.

| Interval | Power reduction of Execution | Requests |
|---|---|---|
| SE6 | 6MW | $f_t^{PCC} \leq 49.60$ and $df_t^{PCC}/dt \leq -11.5$ |
|  |  | $f_t^{PCC} \geq 50.40$ and $df_t^{PCC}/dt \geq 11.5$ |
| SE5 | 5MW | $49.60 \leq f_t^{PCC} < 49.65$ and $-11.5 < df_t^{PCC}/dt \leq -10$ |
|  |  | $50.35 \leq f_t^{PCC} < 50.40$ and $10 \leq df_t^{PCC}/dt < 11.5$ |
| SE4 | 4MW | $49.70 \leq f_t^{PCC} < 49.65$ and $-10.0 < df_t^{PCC}/dt \leq -8.0$ |
|  |  | $50.30 \leq f_t^{PCC} < 50.35$ and $8.0 \leq df_t^{PCC}/dt < 10.0$ |
| SE3 | 3MW | $49.75 \leq f_t^{PCC} < 49.70$ and $-8.0 < df_t^{PCC}/dt \leq -6.5$ |
|  |  | $50.25 \leq f_t^{PCC} < 50.30$ and $6.5 \leq df_t^{PCC}/dt < 8.0$ |
| SE2 | 2MW | $49.80 \leq f_t^{\text{PCC}} < 49.75$ and $-6.0 < \mathrm{d}f_t^{\text{PCC}}/\mathrm{d}t \leq -4.5$ |
| | | $50.20 \leq f_t^{\text{PCC}} < 50.25$ and $4.5 \leq \mathrm{d}f_t^{\text{PCC}}/\mathrm{d}t < 6.0$ |
| SE1 | 1MW | $49.80 \leq f_t^{\text{PCC}} < 49.75$ and $-4.5 < \mathrm{d}f_t^{\text{PCC}}/\mathrm{d}t \leq -2.5$ |
| | | $50.20 \leq f_t^{\text{PCC}} < 50.25$ and $2.5 \leq \mathrm{d}f_t^{\text{PCC}}/\mathrm{d}t < 4.5$ |

## 5 Battery optimization

Based on the proposed full-time scale EMS, the production simulation over a reference year (8760 hours) is performed. The battery size optimization model is established with an objective of the minimum LCOH and constraints of system operation stability and the long-term energy balance. Then, an iterative searching algorithm based on the off-line simulation is developed to assess the most cost-effective battery configuration for a real OReP2HS.

*5.1 Objective*

The optimization objective set to minimize the LCOH, written as:

$$\min LCOH = \min \frac{(C_{\text{inve}} + C_{\text{O\&M}}^{\text{fixed}} + C_{\text{O\&M}}^{\text{vari}})}{M^{\text{H}_2}} \tag{23}$$

where, the cost of hydrogen in the OReP2HS consist of the initial investment cost $C_{\text{inve}}$, fixed annual operation and maintenance (O&M) cost $C_{\text{O\&M}}^{\text{fixed}}$, and variable annual O&M cost $C_{\text{O\&M}}^{\text{vari}}$. $M^{\text{H}_2}$ is the annual yield of hydrogen. The detailed calculations of the hydrogen production cost are present in (24)-(32).

$$C_{\text{inve}} = \sum_j CRF(r, T_j^{\text{LCC}}) S_j c_j^{\text{unit}} \tag{24}$$

$$CRF(r, T_j^{\text{LCC}}) = \frac{r(1+r)^{T_j^{\text{LCC}}}}{(1+r)^{T_j^{\text{LCC}}} - 1} \tag{25}$$

$$C_{\text{O\&M}}^{\text{fixed}} = \sum_j \lambda_j^{\text{O\&M}} S_j c_j^{\text{unit}} \tag{26}$$

$$C_{\text{O\&M}}^{\text{vari}} = \sum_j (C_j^{\text{rep}} - C_j^{\text{rec}}) \tag{27}$$

$$C_j^{\text{rep}} = \sum_{k=1}^{K_j^{\text{rep}}} S_j^{\text{rep}} c_j^{\text{rep}} (1+r)^{-kT_j^{\text{LCC}}/(K_j^{\text{rep}}+1)} \tag{28}$$

$$C_j^{\text{rec}} = \sum_{k=1}^{K_j^{\text{rep}}} S_j^{\text{rep}} c_j^{\text{rec}} (1+r)^{-kT_j^{\text{LCC}}/(K_j^{\text{rep}}+1)} \tag{29}$$

$$K_j^{\text{rep}} = \left[\frac{T_j^{\text{LCC}}}{T_j^{\text{Fail}}} - 1\right] \tag{30}$$

$$T_j^{\text{Fail}} = \frac{\Delta E_j^{\text{degra},max}}{\Delta E_j^{\text{degra},year}} \tag{31}$$

$$M^{\text{H}_2} = \sum_t \sum_i (\alpha_{\text{P2H}}(q_t^{\text{H}_2}, P_t^{\text{AE}}) P_{i,t}^{\text{AE}} \Delta t) \tag{32}$$

(24)-(25) describe the initial cost, where $S_j$ is the device capacity, $c_j^{\text{unit}}$ is the unit cost of devices. $CRF(r, T_j^{\text{LCC}})$ is the capital recovery factor. $r$ is a discount rate, and it is set at 8%. $T_j^{\text{LCC}}$ is the lifetime of the project. (26) assumes the fixed annual O&M cost $C_{\text{O\&M}}^{\text{fixed}}$ as a fixed proportion of the initial investment cost, $\lambda_j^{\text{O\&M}}$ is the ratio coefficient. (27)-(29) describe the variable annual O&M cost as the battery replacement cost $C_j^{\text{rep}}$ and benefit of the retired battery recycling $C_j^{\text{rec}}$, where $S_j^{\text{rep}}$ is the replaced battery capacity, $c_j^{\text{rep}}$ and $c_j^{\text{rec}}$ unit cost and benefit of the battery replacement and recovery, respectively. (30)-(31) present the calculation of represents the times of battery replacement over the whole project lifetime. $\Delta E_j^{\text{degra},year}$ is annual degradation, and its detailed calculation can be seen in [46] [47]. $\Delta E_j^{\text{degra},max}$ is the maximum degradation of a healthy battery, and it typically takes 20%. The annual hydrogen production is described in (32), where $\alpha_{\text{P2H}}(q_t^{\text{H}_2}, P_t^{\text{AE}})$ is the conversion efficiency function extracted from a practical AE.

*5.2 Constrains*

Three requirements for the OReP2HS design mentioned in Section 2 should be checked in the battery size optimization. It includes GFM ability, continuous operation ability during emergencies, and long-term energy balance. However, the GFM ability and continuous operation ability during an emergency are hard to explicitly express and quantify. Thus, based on the refined and simplified simulation models, an off-line simulation is employed to check the stability constraints. Also, the frequency and bus voltage fluctuation at the point of common connection (PCC) of the simulation are set to meet the constraints in (33).

$$\begin{cases} f_{min}^{PCC} \leq f_{tsim}^{PCC} \leq f_{max}^{PCC} \\ V_{min}^{PCC} \leq V_{tsim}^{PCC} \leq V_{max}^{PCC} \end{cases} \quad (33)$$

where, $f_{min}^{PCC}$ and $f_{max}^{PCC}$ are the allowable boundaries of the frequency varication, $V_{min}^{PCC}$ and $V_{max}^{PCC}$ are the allowable boundaries of the bus voltage varication. $f_{tsim}^{PCC}$ and $V_{tsim}^{PCC}$ is the simulation frequency and bus voltage at each simulation step, respectively.

*5.3 Iterative searching based solving algorithm*

An off-line simulation-based iterative searching algorithm is developed for solving the battery size optimization. The framework of the main process of the proposed iterative searching algorithm is illustrated in Fig. 4.

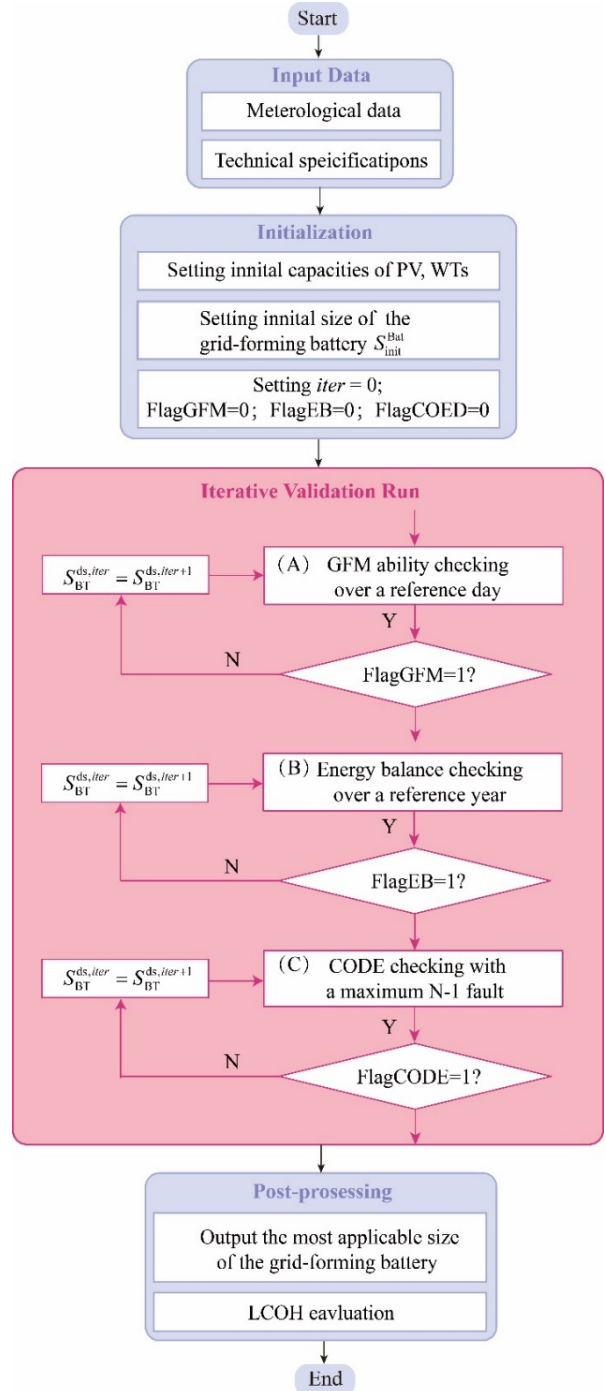

**Fig. 4** Framework of main process of the proposed iterative searching algorithm.

The first step of the iterative searching algorithm requires an initial capacity of battery to start the simulation. It can be set as 5% of the total installed capacity of the power sources with a high charging/discharging rate (2C or 3C), according to the engineering experience. To facilitate the subsequent searching, a preset configuration table for battery capacity iterative search is listed as follows:

Table. 3 Iteratively searching preset configuration table.

| Preset configuration | | | | | | |
|---|---|---|---|---|---|---|
| Iterative variable | $S_{BT}^{ds,0}$ | $S_{BT}^{ds,1}$ | $S_{BT}^{ds,2}$ | ... | $S_{BT}^{ds,iter-1}$ | $S_{BT}^{ds,iter}$ |
| Value | $(S_{init}^{Bat}, 2c)$ | $(S_{init}^{Bat} + \Delta S^{Bat}, 2c)$ | $(S_{init}^{Bat} + \Delta S^{Bat}, 3c)$ | ... | $(S_{init}^{Bat} + int(iter/2)\Delta S^{Bat}, 2c)$ | $(S_{init}^{Bat} + (iter/2)\Delta S^{Bat}, 3c)$ |

$iter \in N$

where $\Delta S_{ES}$ is the increment of battery capacity, $int(\cdot)$ is a ceiling function.

After the preset and initiation, the iterative searching begins. Three subroutines simulate and verify the constraints of GFM ability, continuous operation ability during emergency, and long-term energy balance in sequence. Once a subroutine fails the verification, the battery capacity is corrected by the preset configuration table and then enters the subroutine again until the verification passes. After all the subroutines are verified, the battery configuration is output, and the LCOH is calculated based on the production simulation over 8760 hours.

The features of each subroutine are different. For the GFM ability checking, the daily operation of the OReP2HS is simulated based on the refined model and the proposed EMS. The energy balance checking employs the simplified model to conduct a production simulation over a reference year. For the CODE checking, the maximum N-1 fault of both the source and load of the OReP2HS tasted on the refined model. Detailed flowcharts of the corresponding subroutine are shown in Appendix B.

## 6 Results and discussion

To assess the feasibility and the superior performance of the proposed method, the practical OReP2HS located in Inner Mongolia, China as introduced in Section 2 is simulated and optimized by means of the developed models in Matlab/Simulink software. The main technical and economic parameters of the generations and loads are listed in Table. 3. The initial data of the battery size iterative optimization are listed in Table. 4. The input meteorological data used for the simulation are shown in Fig. 5. Here, the observed and forecasted wind speed and solar irradiation data sets with 1s~60s time resolution are collected from the wind farm and PV station, respectively. It should be noted that the input data with 1s time resolution is used for supporting the simulation of the GFM ability checking and CODE checking, while the energy balance checking over a full year is performed based on the data with 60s resolution. The ambient temperature is obtained by the POWER Data Access Viewer, and its time resolution is only 1h [48].

Table. 3 Main techno-economic parameters for the OReP2HS.

| Facilities | Number | Rated power | Unit investment cost | O&M proportionality coefficient | Project lifetime |
|---|---|---|---|---|---|
| WT | 3 | 6250 kW | 5000 CYN/kW | 2% | 20 years |
| PV | 1 | 5000 kW | 4000 CYN/kW | 2% | 20 years |
| AE | 4 | 5000 kW | 3500 CYN/kW | 2% | 20 years |
| Battery | 1 | Optimized | 1500 CYN/kWh | 2% | 20 years |

Table. 4 Initial data of the battery size iterative optimization.

| Parameter | Value | Parameter | Value |
|---|---|---|---|
| $T_{ROSE}$ | 4 h | $SOC_{max}^{Bat}$ | 0.0142 |
| $C^{H_2}$ | 30 CYN/kg | $\beta$ | 0.0286 |
| $C_{up,h}^{AE}$ | 2 CYN | $c_j^{rep}$ | 900 CYN/kWh |
| $C_{up,c}^{AE}$ | 10 CYN | $c_j^{rec}$ | 150 CYN/kWh |
| $C_{down}^{AE}$ | 5 CYN | $f_{min}^{PCC}$ | 45 Hz |
| $s$ | 1000 CYN/MW | $f_{max}^{PCC}$ | 55 Hz |
| $T_{min}^{down}$ | 1 h | $V_{min}^{PCC}$ | 31.5 kV |
| $k_i^{H_2}$ | 55.62 kWh/kg | $V_{max}^{PCC}$ | 38.5 kV |
| $SOC_{min}^{Bat}$ | 10% | $\Delta S_{ES}$ | 0.1 MWh |
| $SOC_{max}^{Bat}$ | 90% | | |

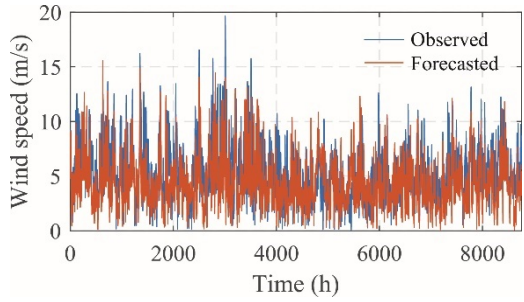

(a) Wind speed

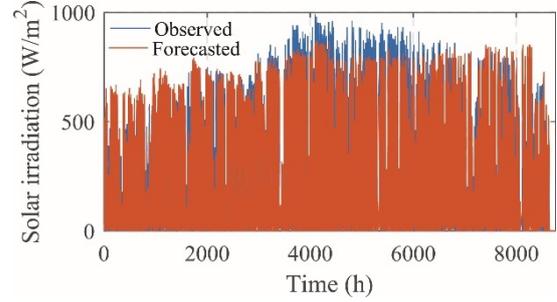

(b) Solar irradiation

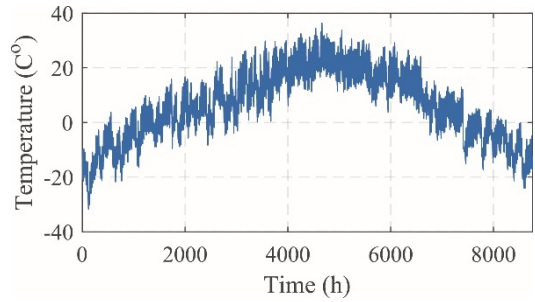

(a) ambient temperature

**Fig. 5** Input meteorological data for the simulation of the practical OReP2HS located in Inner Mongolia, China

*6.1 Simulation and the optimization results*

6.1.1 Optimal control and battery size

Based on the proposed full-time scale EMS, optimization is performed by the iterative searching algorithm to identify the battery size for supporting the stability operation of the OReP2HS. The results are listed in Table. 5 shows the most suitable configuration of the battery is 6.8MW/3.4MWh, which means the battery can be fully charged within 0.5 hour at the rated power, and it provides a fast power support of twice the rated capability. Here, the optimized battery size indicates that, as a GFM source of the OReP2HS, the fast power-supporting ability of the battery is more important than its long-term energy transfer capability. It is consistent with the role and action of a GFM battery designed in the proposed full-time scale EMS. On the other hand, beneficial from the SLF, the load regulations of AEs replace the battery to maintain the energy balance beyond 4s and release the energy transfer requirement from the battery. Thus, the optimized capacity of the battery (3.4MWh) only accounts for 13.6% of the total installed capacity of the sources. The detailed model introduced in Section 2 is the basis to support the simulation with such a high precision.

**Table. 5** key operation parameters and the corresponding optimization results.

| Time step of the SLF | Ramp rate of AEs | Battery size | Charge-/Discharge-Rate | Yearly degradation | Maximum AE's load regulation within a time step of SLF | LCOH |
|---|---|---|---|---|---|---|
| 5 s | 0.05 MW/s | 3.4 MWh | 2C | 4.87 % | 0.25 MW | 28.829 CYN/kg |

The production simulation (energy balance simulation) of the OReP2HS over a reference year is illustrated in Fig.6. And the simulation results from October 1st to October 7th are shown in detail. It can be seen that, with the SLF strategy, the AEs can follow the renewable power fluctuations closely, which keeps an efficient utilization of renewable energy and increases hydrogen production. During the periods of 6514h–6524h, 6539h–6548h, and 6608h–6619h, the outputs of WTs and PV are complementary, which avoids energy waste of the shut-down and -up of the AEs. It shows that equipping both WTs and PV generations to produce hydrogen is beneficial, AEs only shun-down for 17.5h during seven days. Furthermore, during the periods of 6596h–6606h, there is only WTs are working and the AEs can match the wind power fluctuations well. However, during the periods of 6608h–6619h, the AEs cannot track the PV power closely due to the power ramp of the PV station being significantly faster than the AE. Thus, the imbalance of power is smoothed by the GFM battery.

Fig.7 shows the SOC and power changes of the GFM battery from Oct. 1st to 7th. It can be observed from Fig.7, with the proposed EMS, that the active power of the battery rapidly changes when WTs working at the long-term high wind speed periods. In contrast, the reactive power changes violently when the PV station working in the daytime. From the holistic perspective, the active power changes more frequently than reactive power, which reflects that maintaining the active power balance (or the frequency stability) is crucial for the stabilized operation of the OReP2HS. The SOC, always keeps within the allowable interval (10%~90%), demonstrating the effectiveness of the SOC linear rectify strategy. In addition, the changing of the SOC is similar to the renewable power, implying a temporal correlation between them, and a more detailed discussion is presented in Section 6.5. The frequency and voltage variation of the point of common connection are plotted in Fig.8. Compared with Fig.7, we can see that the variation tendency of the frequency and voltage are accordant with the active and reactive power of the battery, respectively. This is determined by the inherent nature of the P-f and V-Q droop control of the GFM battery. The key technical indexes are listed in the Table. 6. It shows the frequency and volvariationriate within the acceptable range during the whole period, which validates the proposed method and maintains the stable operation of the system well.

In short, the production simulation results show that, with the proposed full-time scale EMS, the OReP2HS operates stably over a reference year with a GFM battery configurated with 6.8MW/3.4M.

Table. 6 Key technical indexes for the operation of OReP2HS.

| Index | Rated value | Allowable interval | Peak value | Nadir value | Average value | Ratio of the maximum deviation to the rating |
|---|---|---|---|---|---|---|
| Frequency | 50 Hz | (45 Hz, 55 Hz) | 50.22 Hz | 49.77 Hz | 50.02 Hz | 0.46% |
| Voltage | 35 kV | (31.5 kV, 38.5 kV) | 36.00 kV | 33.01 kV | 34.89 kV | 5.69% |

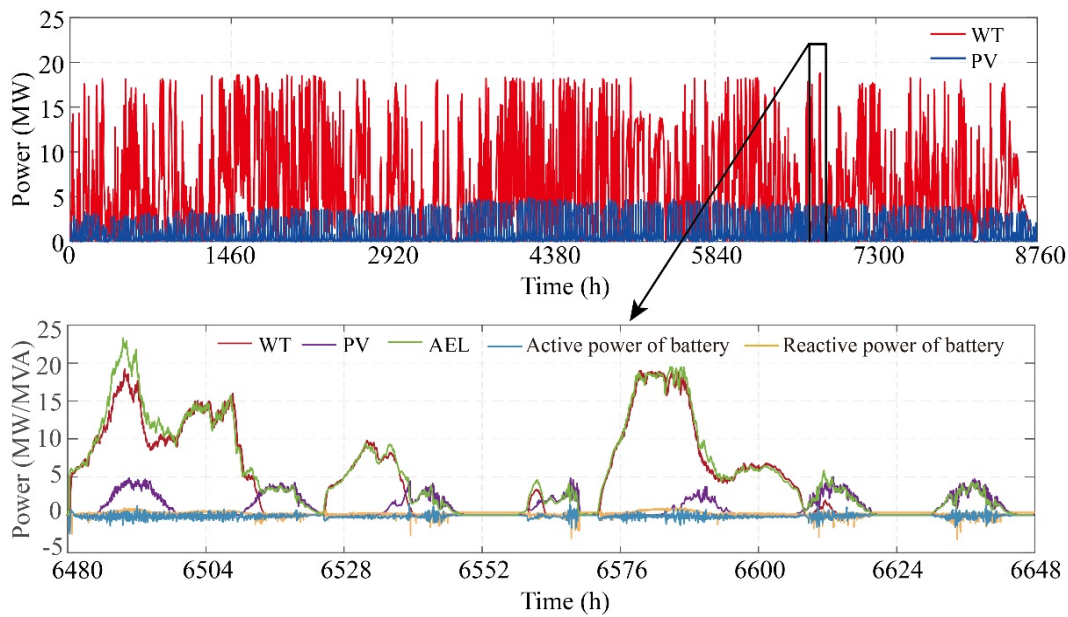

**Fig. 6** Full year production simulation results of the renewable power outputs and load variation of AEs

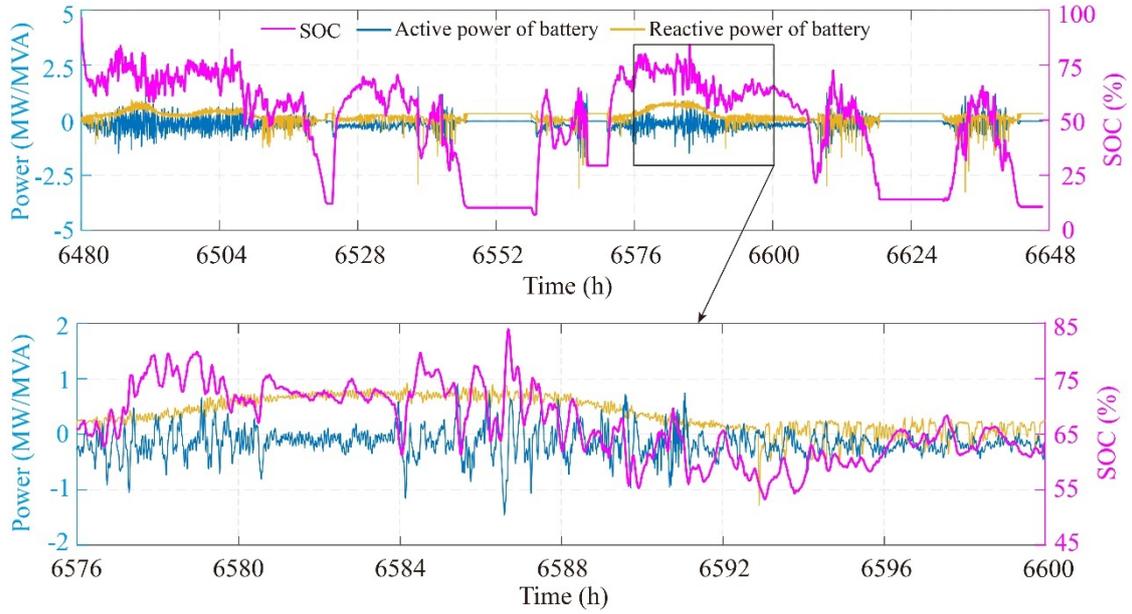

**Fig. 7** Variations of the SOC and power changes of the GFM battery from the Oct. 1st to 7th.

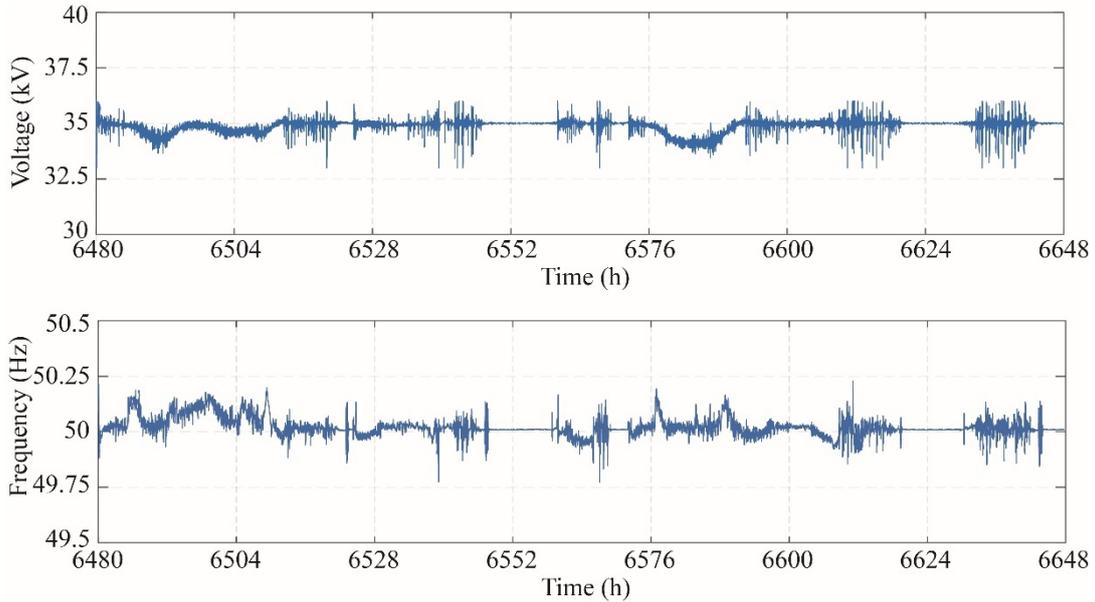

**Fig. 8** The frequency and voltage variation of the point of common connection from the Oct. 1st to 7th.

### 6.1.2 Operation during the emergency

To analyze the performance of the proposed SoCODE, a WT drop test of the OReP2HS with the optimized battery configuration of 6.8MW/3.4MWh is conducted. Both the following scenarios are taken as a contrast.
- Scenario A: the SoCODE is set to be automatically activated based on its detection.
- Scenario B: the SoCODE is removed.

Fig. 9 illustrates the simulation results of power variation of the OReP2HS with a WT dropping at 5s. The corresponding system frequency and battery SOC are depicted in Fig. 10. It can be seen from both Fig. 9 and Fig. 10 that when the emergency occurs at 5s, the frequency drops sharply, while the battery output increases rapidly to avoid the occurrence of frequency collapse. The SoCODE is activated when the frequency deviation reaches 0.4 Hz. Then, the power reduction command is allocated to all the AEs. The load decreases from 15 MW to 5 MW with a ramp of 0.5MW/s in 10 s, and the battery output reduces to the initial state in 5 s. The frequency recovered to the rating within 5s but overregulated to around 50.2 Hz and kept it untiled the SoCODE ended. In contrast, without the SoCODE, the AEs cannot respond to the emergency imminently. Thus, the loads slowly change from 15 MW to 5 MW in 10 s, and the frequency increases tardily. During this period, the battery keeps the power output, its SOC decreases by 10% within 10 seconds.

It should be pointed out that both scenarios (with and without the SoCODE) keep the system stable when the emergency occurs, it indicates the optimized size of the battery satisfies the requirement for supporting OReP2HS to maintain continuous operation. The foremost aim of the proposed SoOCDE is utilizing the flexibility of AEs to decrease the SOC changing during the emergency, so as to save the battery capacity. Furthermore, the execution of the proposed SoCODE is designed reasonably but not perfectly, it still needs to be improved. It is also necessary to deeply exploit the potential and the limitations of the OReP2HS to address emergencies.

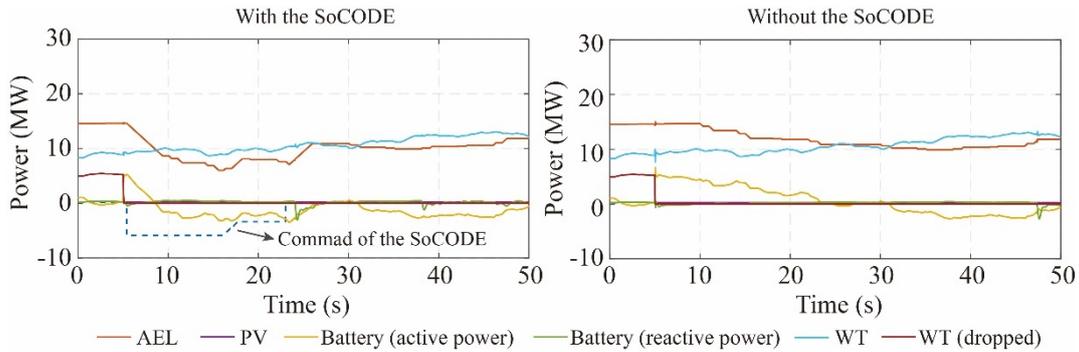

**Fig. 9** System power variations of the OReP2HS when a WT drooped.

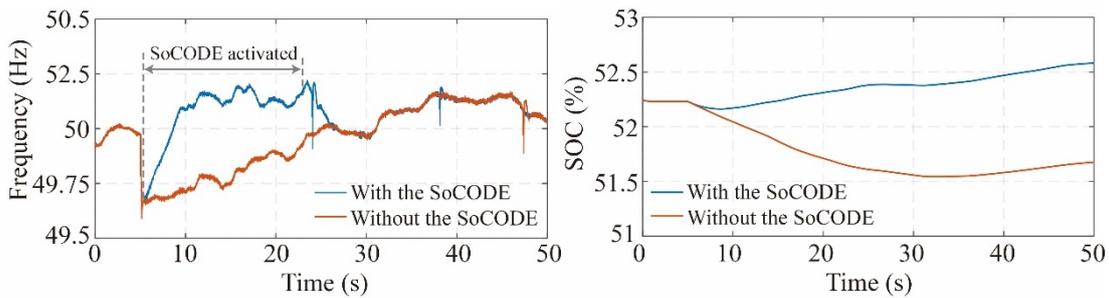

**Fig. 10** The frequency and SOC variation when a WT drooped.

6.1.4 Seasonal hydrogen production

To analyze the hydrogen production of the OReP2HS with the proposed full-time scale EMS, the 12 months of a year are divided into four seasons: spring (Mar. Apr. May), summer (Jun. Jul. Aug.), autumn (Sep. Oct. Nov.) and winter (Dec. Jan. Feb.). The daily hydrogen production with 1h time resolution of each season is depicted in Fig. 11. Due to the AEs' loads tracking the renewable power closely with the proposed SLF strategy (time step is 4s), thus, from a long-term perspective, the hydrogen yield is also influenced by the seasonal meteorological conditions. The daily hydrogen production in winter (2032.32 kg/day on average) is much less than that in the other three seasons, on account of the lower wind speed and the weaker irradiation. From a short-term perspective, the hydrogen production from 12 o'clock to 18 o'clock is mostly the largest in a day, which is affected by the operation feature of the PV station. In contrast, the production stops after sunset or at midnight sometimes.

The monthly average daily production is illustrated in Fig. 12, and the maximum rate of hydrogen production variation within minutes of each month is listed in Table. 7. As the results show, the average daily production fell from 3465.20 kg/day to 1723.47 kg/day in Mar. to May, but increased to 3032.51 kg/ day in Jul. The maximum change rate of hydrogen is 9.90 kg/min, occurring in January. Both the above two indexes are instructive for the practical cases. The former can be used for guiding the production planning of the hydrogen consumers, and the latter is significant for the designing of the hydrogen buffer tanks.

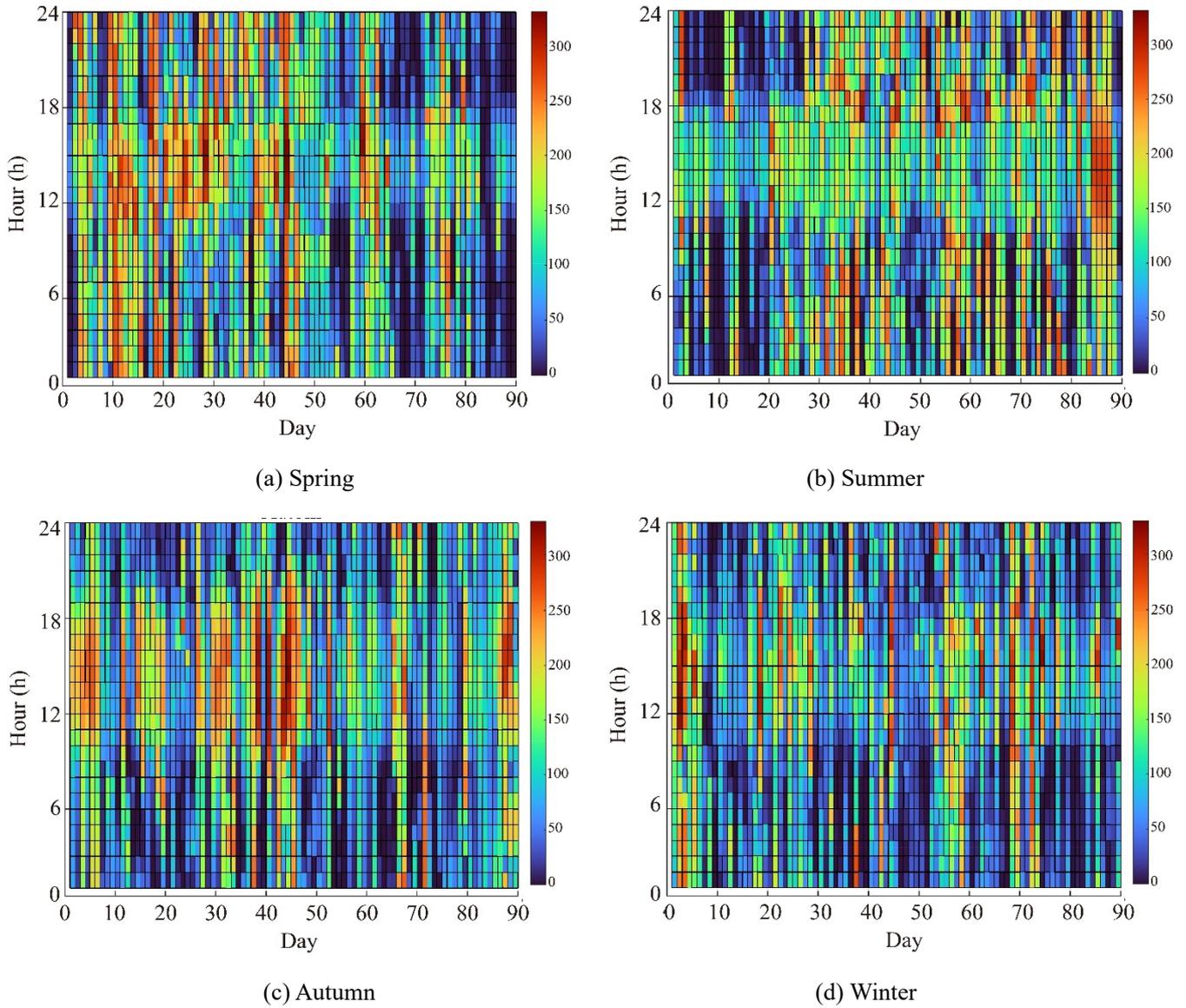

(a) Spring  (b) Summer

(c) Autumn  (d) Winter

**Fig. 11** daily hydrogen productions with 1h time resolution of each season.

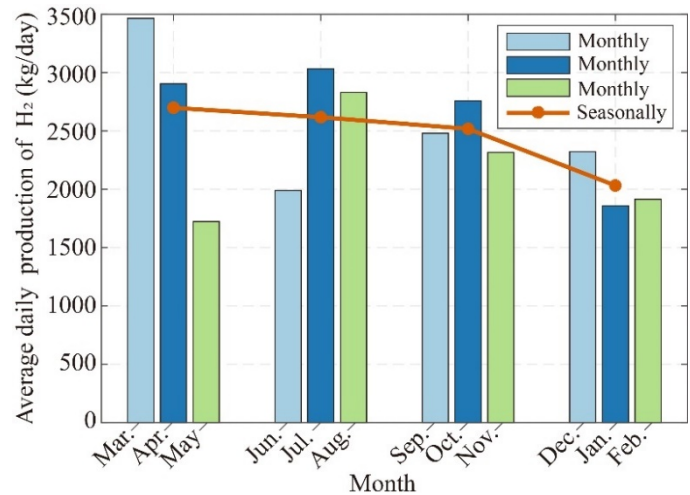

**Fig. 12** Monthly average daily hydrogen production.

**Table. 7** The maximum rate of hydrogen production variation within minutes of each month.

| | Mar. | Apr. | May | Jun. | Jul. | Aug. | Sep. | Oct. | Nov. | Dec. | Jan. | Feb. |
|---|---|---|---|---|---|---|---|---|---|---|---|---|
| Changing rate (kg/min) | 9.50 | 6.05 | 7.85 | 5.56 | 9.77 | 8.44 | 8.97 | 8.43 | 8.11 | 8.64 | 9.90 | 8.97 |

6.1.5 Correlation between the battery SOC and renewable outputs

As we observed from Fig. 7, the changing of the battery's SOC is similar to the renewable power's. Thus, a further inquiry of the correlation between them is analyzed, to reveal the operation feature of the GFM battery with the proposed EMS.

Fig.13 shows the temporary changing of local extremums of the battery SOC and renewable power in Mar. As shown, when the renewable power reaches its local extremum, it is always accompanied by the appearance of a partial peak in the SOC changing. The time of the partial peak appearance of SOC slightly lags behind that of renewable power. It indicates that there is a temporary correlation between the fluctuation of the SOC and renewable power. Then, a statistical analysis of the yearly SOC and renewable output with 1-minute time resolution is performed, and the results are illustrated in Fig. 14. As it is visually shown in Fig. 14 (a), the changing of the SOC shows a positive and nonlinear correlation with the renewable output variation. The SOC changes fast with the renewable output variation in the low output interval (altitude change from 0~0.4 of the normalized power) but gradually slowly after beyond it. Specifically, the battery smooths the imbalance of power between the loads and sources, its SOC is related to the precision of the fast forecasting integrated into the SLF strategy. During the low output interval of renewable outputs, enduring ramps of the renewable generation happen frequently, which increases the difficulty of accurate forecasting. Thus, the battery needs to frequently adjust its charging or discharging states as it is shown in the local statistical results of Fig. 14 (b), and there is almost a linear relationship between the SOC and renewable power. However, after the renewable generations finish ramping and work at a high output interval, their power outputs generally tend to be in a relatively stable state. Now, precise forecasting is much easier than before, so the power could almost be consumed by the AEs, the battery only has to balance the small power jitter. Therefore, the SOC exhibits an irrelevance with the power variation at these periods. In addition, Fig. 14 (b) shows the distributions of the normalized SOC are mainly concentrated around two areas with amplitudes of 0.15 and 0.75, respectively. It reflects the battery mostly works at a deep charge or discharge state, that's why the battery capacity decays rapidly by 4.87% in one year.

From a numerical perspective, Table. 8 demonstrates there is certainly a strong positive correlation between SOC and renewable power by the rank correlation coefficient. Based on this, the SOC change can be predicted according to the forecasting outcomes of renewable power. And actions can be taken in advance to prevent the aging of the battery and optimize the battery management, or even to further reduce the battery size.

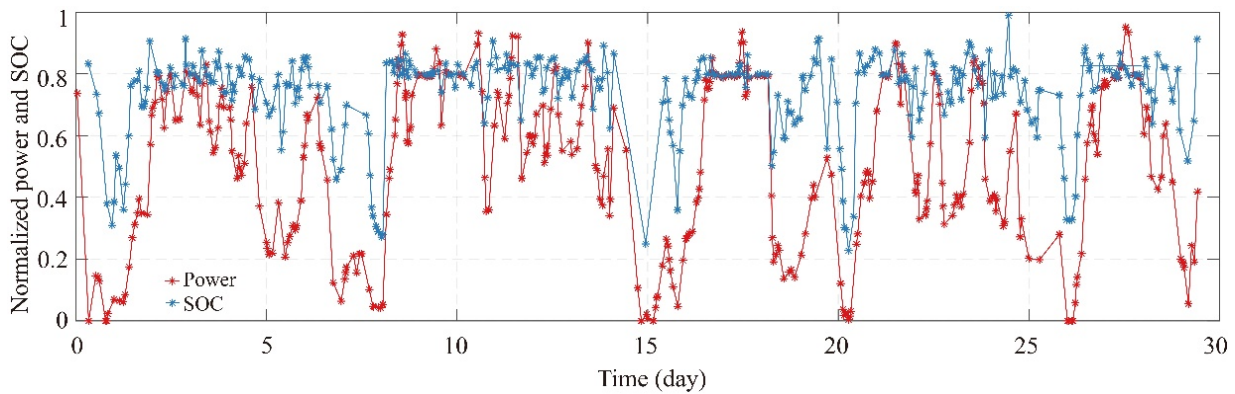

**Fig. 13** The temporary changing of the local extremums of battery SOC and renewable power in March.

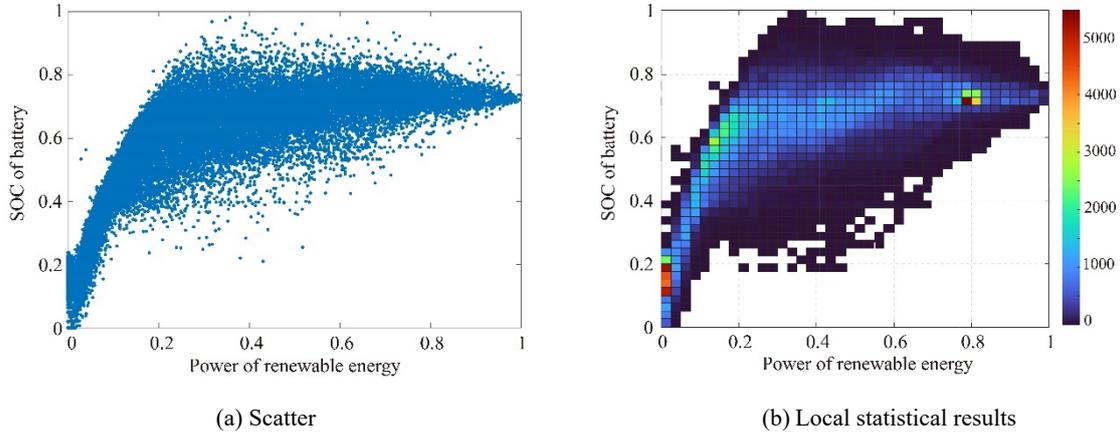

(a) Scatter  (b) Local statistical results

**Fig. 14** Statistical results of the battery SOC and renewable power.

**Table. 8** The rank correlation coefficient between SOC and renewable power in each month.

|  | Mar. | Apr. | May | Jun. | Jul. | Aug. | Sep. | Oct. | Nov. | Dec. | Jan. | Feb. |
|---|---|---|---|---|---|---|---|---|---|---|---|---|
| Rank correlation coefficient | 0.855 | 0.851 | 0.871 | 0.902 | 0.870 | 0.835 | 0.833 | 0.858 | 0.845 | 0.896 | 0.858 | 0.847 |

## 6.2 Sensitivity analysis

Taking advantage of the load flexibility of AEs to replace part of the energy balance regulation for the GFM battery, so as to reduce the battery size and LCOH is the core of the proposed full-time scale EMS. The load flexibility of AEs is significantly influenced by two key adjustable parameters: the time step of the SLF and the ramp rate of AE. Thus, a sensitivity analysis over these two parameters is performed to provide a comprehensive understanding of the limitations of the proposed EMS and show the potential to reduce the LCOH. Specifically, this analysis is executed by varying individual parameters with another one remaining its reference values, and the simulation results are listed in Table 9.

**Table. 9** Simulation results of the sensitive analysis.

| Scenarios | Time step of the SLF | Ramp rate of AEs (MW/s) | Battery size (MWh) /Charge-rate | Yearly degradation | Maximum AE's load regulation within a time step of SLF(MW) | LCOH (CYN/kg) |
|---|---|---|---|---|---|---|
| **Base** | **5s** | **0.05** | **3.40 / 2C** | **4.87%** | **0.250** | **28.829** |
| I | 10s | 0.05 | 3.75 / 2C | 4.88% | 0.500 | 29.798 |
|  | 15s | 0.05 | 4.04 / 2C | 4.91% | 0.750 | 30.805 |
|  | 30s | 0.05 | 4.62 / 2C | 5.01% | 1.443 | 33.099 |
|  | 60s | 0.05 | 5.07 / 2C | 4.97% | 1.888 | 35.551 |
|  | 90s | 0.05 | 5.73 / 2C | 4.99% | 2.334 | 37.814 |
| II | 5s | 0.1 | 3.05 / 2C | 4.94% | 0.500 | 26.731 |
|  | 10s | 0.1 | 3.18 / 2C | 4.94% | 0.933 | 27.212 |
|  | 15s | 0.1 | 3.56 / 2C | 5.04% | 1.214 | 28.497 |
|  | 30s | 0.1 | 3.96 / 2C | 5.05% | 1.443 | 29.898 |
|  | 60s | 0.1 | 4.14 / 2C | 5.04% | 1.888 | 30.891 |
|  | 90s | 0.1 | 4.87 / 2C | 5.12% | 2.334 | 33.094 |
| III | 5s | 0.2 | 2.78 / 2C | 4.94% | 0.710 | 25.836 |
|  | 10s | 0.2 | 2.89 / 2C | 4.93% | 0.933 | 26.246 |
|  | 15s | 0.2 | 3.24 / 2C | 5.03% | 1.214 | 27.417 |
|  | 30s | 0.2 | 3.60 / 2C | 5.11% | 1.443 | 28.663 |
|  | 60s | 0.2 | 3.76 / 2C | 5.12% | 1.443 | 29.520 |
|  | 90s | 0.2 | 4.32 / 2C | 5.14% | 2.334 | 31.091 |

| | 5s | 0.3 | 2.66 / 2C | 4.92% | 0.710 | 25.458 |
| --- | --- | --- | --- | --- | --- | --- |
| | 10s | 0.3 | 2.82 / 2C | 4.93% | 0.933 | 26.024 |
| IV | 15s | 0.3 | 3.14 / 2C | 5.05% | 1.214 | 27.077 |
| | 30s | 0.3 | 3.49 / 2C | 5.13% | 1.443 | 28.286 |
| | 60s | 0.3 | 3.64 / 2C | 5.14% | 1.888 | 29.084 |
| | 90s | 0.3 | 4.15 / 2C | 5.14% | 2.334 | 30.479 |
| | 5s | 0.4 | 2.66 / 2C | 4.97% | 0.710 | 25.453 |
| | 10s | 0.4 | 2.80 / 2C | 4.97% | 0.933 | 25.956 |
| V | 15s | 0.4 | 3.10 / 2C | 5.10% | 1.214 | 26.942 |
| | 30s | 0.4 | 3.43 / 2C | 5.11% | 1.443 | 28.081 |
| | 60s | 0.4 | 3.59 / 2C | 5.15% | 1.888 | 28.903 |
| | 90s | 0.4 | 4.07 / 2C | 5.16% | 2.334 | 30.194 |
| | 5s | 0.5 | 2.66 / 2C | 3.50% | 0.710 | 25.451 |
| | 10s | 0.5 | 2.78 / 2C | 4.97% | 0.933 | 25.891 |
| VI | 15s | 0.5 | 3.08 / 2C | 5.10% | 1.214 | 26.875 |
| | 30s | 0.5 | 3.39 / 2C | 5.11% | 1.443 | 27.945 |
| | 60s | 0.5 | 3.56 / 2C | 5.15% | 1.888 | 28.787 |
| | 90s | 0.5 | 4.02 / 2C | 5.16% | 2.334 | 30.017 |

6.2.1 Battery size

In terms of the optimal battery capacity, results from the sensitive analysis are shown in Fig. 15 and Fig. 16. It can be observed that the battery capacity increases almost exponentially once the ramp rate is lower than 0.3 MW/s in all scenarios. For the higher ramp rate which is beyond 0.3 MW/s, the battery capacity remains at its former level with a slight reduction of 2% in most scenarios. It indicates that the ramp rate of 0.3MW/s of AEs is an inflection point for the optimization of battery size. However, the reduction of the battery capacity caused by the speed-up of the load ramp is decreasing with a longer step of SLF. It implies the coupling of the influence on the battery size optimization caused by these two key parameters.

Regarding the influence of the time step of the SLF, reducing the time step tends to decrease the battery capacity. Also, the reduction of battery capacity caused by the unit change of time step is more than that of the ramp rate. Numerically, with the time step and ramp rate changed to ten times the base value, the average reduction of battery capacity is 0.876 MWh and 0.484 MWh, respectively. It shows the time step has a major effect on the battery size optimization. Therefore, with the proposed full-time scale EMS, a shorter time step of the SLF has an edge over a higher ramp rate of AEs.

From the results depicted in Fig. 16, the minimum battery size of 2.66MWh appears when the ramp rate of AEs is 0.5MW/s with a 5-second time step of the SLF. However, load changes with 0.5MW/s may be difficult for the designing of AEs. The ramp rate adopts in the range of 0.2 MW/s to 0.3 MW/s with 5s or 10s time step of the SLF is more advisable, and the battery size changes in an acceptable interval of 2.66MWh to 2.89MWh.

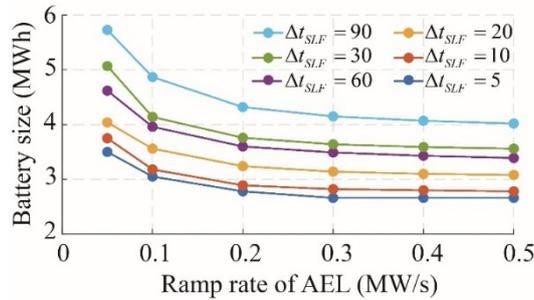

**Fig. 15** Battery size changing result in different scenarios.

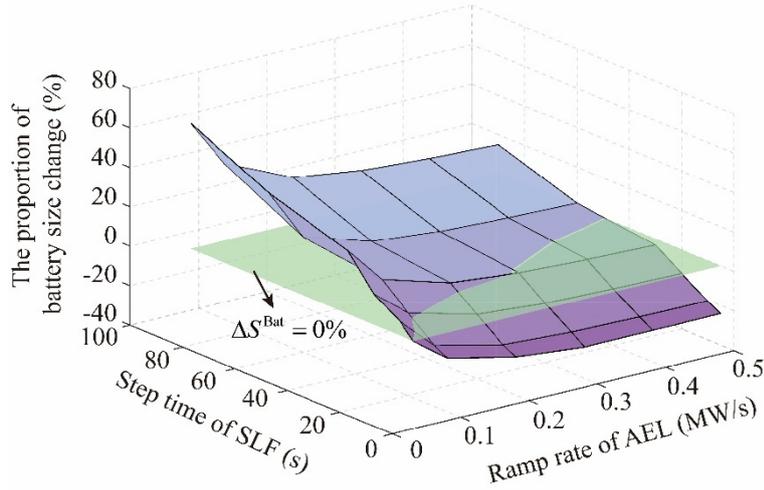

**Fig. 16** The changing proportion of battery based on the base-scenario.

6.2.2 LCOH

The results of the sensitivity analysis on the LCOH are shown in Fig. 17. For the influence of the two key parameters on the LCOH, the cost of hydrogen can be seen to decrease nearly linearly as a function of the time step of SLF. Moreover, the changing rate is significantly faster after the time step is shorter than 30s. The ramp rate of AE has a low impact on the LCOH after it goes beyond 0.3MW/s. Nevertheless, once the ramp rate decreases below 0.3MW/s, the LCOH starts to rise exponentially and reaches its maximum value at 0.05MW/s in all scenarios.

Compared with Fig. 16, it can be seen that LCOH and battery size change similarly. It reflects the battery cost (including initial investment, and fixed and variable O&M cost) has a major effect on the cost of hydrogen production in the studied case. Also, the variation tendency of the LCOH influenced by both the ramp rate and time step of SLF are in keeping with that of the battery. The minimum LCOH of 25.451 CYN/kg appears when the ramp rate of AEs is 0.5MW/s with a 5-second time step of the SLF. And it still has competitive prices in intervals of 26.246 CYN/kg to 25.458 CYN/kg, when the ramp rate is adopted in a load-friendly range of 0.2 MW/s to 0.3 MW/s with 5-second or 10-second time step of the SLF.

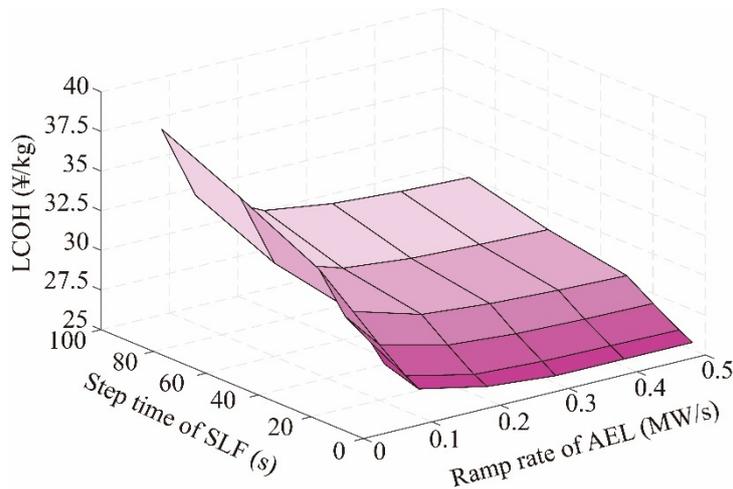

**Fig. 17** LCOH changing result in different scenarios.

6.2.3 Fast load regulation of AE

The precise and fast load regulation of SLF is essential for the battery size and LCOH reduction in the proposed EMS. Fig. 18 shows the change in proportion of the maximum AE's load regulation to rating within a time step of SLF. As we can see, the maximum load regulation of AE keeps stable with the speed op of the ramp rate, when the time step is longer than 30s. However, the average utilization proportion of the fast load regulation space is only 31.1% in these scenarios, showing a huge gap between the actual load regulation and theoretical ability.

In contrast, in the scenarios of a shorter time step of SLF (5s~30s), the average utilization proportion is increased with 36.2% increments. Especially, the fast load regulation ability is fully released when the ramp rate is set to 0.05 MW/s with 5s~15s time step. However, the regulation ability cannot fully satisfy the energy balance requirement, and the load flexibility is also limited. Until the ramp rate accelerates to 0.2 MW/s or 0.3 MW/s, the load regulation remains stable but not saturation, showing the energy balance requirement can be matched well.

Therefore, comprehensively, the ramp rate is adopted in a load-friendly range of 0.2 MW/s to 0.3 MW/s with 5-second or 10-second time step of the SLF is worth recommending in the studied case.

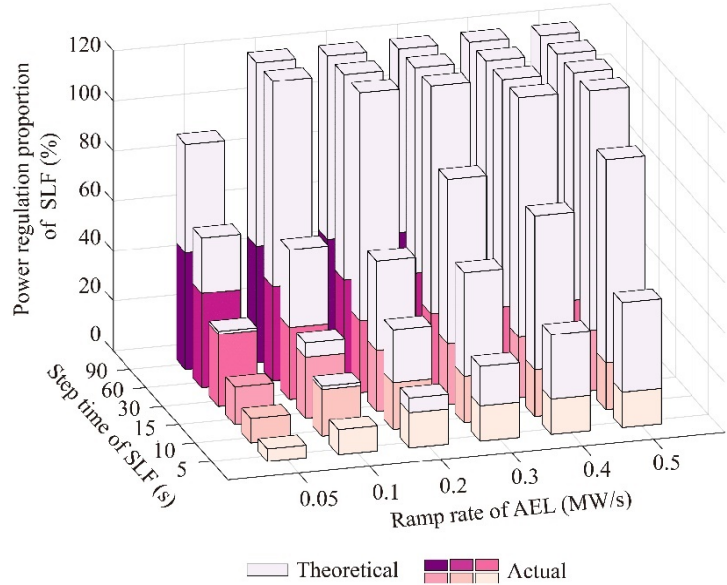

**Fig. 18** The proportion of the maximum load regulation of AE to rating within a time step of SLF.

## 7. Conclusion

A full-time scale EMS that takes battery as a GFM source of an OReP2HS is presented, it covers from GFM control to system scheduling (from milliseconds to hours). Then, the battery size is identified based on the EMS, while considering the constraints of the GFM ability, continuous operation ability during emergency, and long-term energy balance.

A practical OReP2H project configurates 18.75 MW WTs, 6.25 MWp PV, and 4 AEs with each rated power of 5 MW located in Inner Mongolian, China is taken as a case study. Simulation results show that the most suitable configuration of the battery is 6.8MW/3.4MWh. Here, the optimized battery size indicates that, as a GFM source, the fast power-supporting ability of the battery is more important than its long-term energy transfer capability. On the other hand, beneficial from the SLF, the load regulation of AEs replaces the battery to maintain the energy balance beyond 4s, releasing the energy transfer requirement from the battery. Thus, the optimized capacity of the battery (3.4MWh) only accounts for 13.6% of the total installed capacity of the sources.

The minimum LCOH is 28.829 CYN/kg with the configuration of the optimum battery size. The hydrogen yield is influenced by the seasonal meteorological conditions, due to the AEs' loads tracking the renewable power closely with the proposed SLF strategy. The average daily production in March approaches 3465.20 kg/day, and the maximum change rate of hydrogen is 9.90 kg/min, occurring in January. Moreover, we find that the changing of the SOC shows a positive correlation with the renewable output variation. It implies that SOC change can be predicted according to the forecasting outcomes of renewable power. Actions can be taken in advance to prevent the aging of the battery and optimize the battery management, or even to further reduce the battery size, these would be interesting areas for future research.

Sensitivity analysis shows the optimized battery size decreases when the increases ramp rate of AE or the time step of SLF. The variation tendency of the LCOH changes similar to battery size, it reflects the battery cost (including initial investment, fixed and variable O&M cost) has a major effect on the cost of hydrogen production. It also reveals the essential to save the battery size for an OReP2HS. Considering from the design and maneuverability of AE comprehensively, the ramp rate is adopted in a load-friendly range of 0.2 MW/s to 0.3 MW/s with 5s or 10s time step of the SLF is worth recommending in the studied case. Here, the configuration of the battery changes from 5.32MW/2.66MWh to 5.78/2.89MWh, and hydrogen has competitive prices in intervals of 26.246 CYN/kg to 25.458 CYN/kg.

☐